\documentclass[pdflatex, sn-mathphys-num]{sn-jnl}

\usepackage{graphicx}%
\usepackage{multirow}%
\usepackage{amsmath,amssymb,amsfonts}%
\usepackage{amsthm}%
\usepackage[scr=rsfso]{mathalpha}   
\usepackage[title]{appendix}%
\usepackage{xcolor}%
\usepackage{xurl}       
\usepackage{microtype}  
\usepackage{textcomp}%
\usepackage{manyfoot}%
\usepackage{booktabs}%
\usepackage{algorithm}%
\usepackage{algorithmicx}%
\usepackage{algpseudocode}%
\usepackage{listings}%
\usepackage{longtable}%
\usepackage{forest}
\usepackage{tikz}
\usepackage{adjustbox}
\usepackage{lmodern}
\usepackage[T1]{fontenc}
\usepackage[utf8]{inputenc}
\usepackage{silence}
\usepackage{subcaption} 
\usepackage{array}        
\usepackage{ragged2e}   
\usepackage{bookmark}
\usepackage{mathpazo}

\newcolumntype{L}[1]{>{\RaggedRight\arraybackslash}p{#1}}
\newcolumntype{C}[1]{>{\Centering\arraybackslash}m{#1}}
\newcolumntype{R}[1]{>{\RaggedLeft\arraybackslash}p{#1}}
\usepackage{makecell}   
\forestset{
  mc/box/.style={
    draw, rounded corners=2pt, align=center,
    inner xsep=4pt, inner ysep=3pt, font=\small
  },
  mc/edge/.style={edge={-latex, line width=0.5pt}},
  mc/base/.style={
    for tree={
      mc/box,
      mc/edge,
      grow=south,     
      anchor=north,
      s sep=4mm,
      l sep=6mm
    }
  },
  mc/one/.style={minimum width=21mm, minimum height=6.5mm},
  mc/two/.style={minimum width=21mm, minimum height=10.5mm},
}

\theoremstyle{thmstyleone}%
\theoremstyle{thmstyletwo}%

\theoremstyle{thmstylethree}%
\definecolor{metric}{RGB}{213, 94, 0}

\newif\ifrevisionnotes
\revisionnotestrue
\ifrevisionnotes
  \newcommand{\hr}[1]{\footnote{\textbf{\textcolor{blue!60!black}{Hindy (revision): #1}}}}
  \newcommand{\gf}[1]{\footnote{\textbf{Giorgio: #1}}}
  \newcommand{\mm}[1]{\footnote{\textbf{Mirco: #1}}}
\else
  \newcommand{\hr}[1]{}
  \newcommand{\gf}[1]{}
  \newcommand{\mm}[1]{}
\fi

\begin{document}

\title{Evaluating AI Musical Creativity: Towards a Quantitative Multi-Dimensional Framework Based on Value, Novelty, and Surprise}

\author[1]{\fnm{Hindy} \sur{Rossignol}}

\author[2]{\fnm{Giorgio} \sur{Franceschelli}}

\author[3,2]{\fnm{Mirco} \sur{Musolesi}}

\affil[1]{\orgdiv{Department of Arts and Sciences}, \orgname{University College London}, \orgaddress{\street{Malet Place}, \city{London}, \postcode{WC1E 7JG}, \country{United Kingdom}}}

\affil[2]{\orgdiv{Department of Computer Science and Engineering}, \orgname{University of Bologna}, \orgaddress{\street{Viale del Risorgimento, 2}, \city{Bologna}, \postcode{40136}, \country{Italy}}}

\affil[3]{\orgdiv{Centre for Artificial Intelligence, Department of Computer Science}, \orgname{University College London}, \orgaddress{\street{66-72 Gower Street}, \city{London}, \postcode{WC1E 6EA}, \country{United Kingdom}}}


\abstract{ 
Despite the increasing sophistication of generative artificial intelligence (AI), research on the evaluation of AI-generated musical creativity remains limited. To address this gap, this paper proposes a quantitative framework for the formalisation and evaluation of AI musical creativity. Building on Margaret Boden’s theory, which conceptualises creativity in terms of value, novelty, and surprise, the framework adapts and operationalises these three dimensions for the musical domain. On the hypothesis that certain aspects of musical creativity can be assessed with a degree of objectivity, a domain-specific set of quantitative metrics for evaluating the creativity of AI-generated music is introduced. Rather than reducing musical creativity to a single unifying measure, this article proposes an ensemble of complementary metrics that capture its different dimensions, recognizing the complexity of musical artifacts and the difficulty of defining meaningful multi-objective compositional criteria. The framework has two implications. First, it gives researchers and developers a principled methodology and quantitative tools for empirically evaluating creative outputs, enabling more informed model development and reducing unnecessary computational costs. Second, it lays the groundwork for future generative models to optimize directly for creativity by adopting these metrics as learning objectives.
}

\maketitle

\section{Introduction}\label{sec1}

Amid the rapid wave of innovation in Artificial Intelligence (AI), music generation has arguably received less attention than text, image, and video generation. Recent advances in these domains include models such as OpenAI's GPT-5 \citep{openai_gpt5} and ChatGPT Images 2.0 \citep{openai_gptimage2}, Google DeepMind's Gemini \citep{googledeepmind_gemini} and Veo \citep{googledeepmind_veo}, and Black Forest Labs' FLUX.2 \citep{blackforest_flux2}. One factor contributing to this disparity is the absence of robust and standardized frameworks for evaluating AI-generated musical artifacts, which makes it difficult to measure progress reliably and to establish clear criteria for comparing generative systems.

AI music generation, as a tool, has the potential to revolutionize tasks such as composition and improvisation, offering substantial value to professionals in creative industries and music consumers alike. For instance, in industries such as movies and advertising, AI-generated music is starting to be used as a quick and cost-effective solution \citep{Totlani_2023}. Similarly, users of platforms like Suno \citep{suno} and Udio \citep{udio} compose songs in their taste from simple descriptive prompts, a practice that has moved from early-adopter curiosity to a sizeable commercial market.

Yet, modern AI systems still appear to fall under the shadow of Ada Lovelace's famous objection: that machines, no matter how sophisticated, can only do what they are told to do---they cannot originate anything truly new \citep{franceschelli_2024_creativity}. This observation resonates with current generative models' outputs, which are often understood to be mimicking existing patterns rather than truly being creative. At its core, teaching machines to be creative is an inherently challenging problem because creativity is a human construct \citep{agres_2016_evaluation}. Two fundamental obstacles hinder progress in this area. First, there is no consensus on a definition of musical creativity in humans, let alone in the context of generative AI. Second, there is no unified framework for objectively quantifying creativity in AI-generated music.

In music schools, for instance, musical creativity is measured through vague mark schemes, in which the definition of creativity resembles that of newness \citep{hickey_1999_assessment}. Assessments rarely specify concrete features or
characteristics that define creative music, leaving significant room for subjective interpretation by the assessor. Therefore, the attempt to construct an evaluation metric for musical creativity begins with a core challenge: how to systematize a concept that is inherently subjective. On one side is the belief that even complex concepts such as creativity can be formalized through general principles. On the other is the skepticism that any such framework can fully capture the deeply intertwined relationship between the artist and their creation. Music is one of the greatest mysteries of humanity, and evaluating it requires drawing on several disciplines at once: listening to and judging music involves a complex interaction between the physical properties of sound, the listener's physiological and neural mechanisms, their sociocultural background, and their personal exposure \citep{hallam_2022_the}. Creativity is no less complex and requires navigating an ocean of frameworks \citep{candy_2013_evaluating}, making it highly impractical to evaluate. This paper investigates the following research questions:

\begin{itemize}
    \item Is it possible to rigorously define AI music creativity?
    \item To what extent can existing frameworks of creativity, such as Margaret Boden's framework of creativity as defined by value, novelty, and surprise \citep{boden_2004_the}, help evaluate musical creativity in AI-generated music?
    \item What metrics could be used to measure the components of musical creativity in AI-generated artifacts?
    \item How objectively can aspects of musical creativity be quantified in artifacts produced by generative models?
\end{itemize}

As a first step towards addressing these questions, definitions of music, creativity, and musical creativity are formalized, leading to a final definition of generative AI musical creativity (M-Creativity) in Section~\ref{sec:mcreativity-definition}. Then, quantitative metrics to evaluate aspects of M-Creativity are developed (Sections~\ref{sec:mvalue}, \ref{sec:msurprise} and \ref{sec:mnovelty}), with the aim of establishing the building blocks for a unified framework to assess creativity in AI-generated music.

To summarize, this paper makes the following contributions. It introduces a definition of machine musical creativity by building on Boden's general creativity framework, going beyond her original work and adapting it to the music domain. It then formalizes a set of quantitative metrics to measure the various components of this definition. Rather than seeking to reduce musical creativity to a single unifying metric, it proposes an \textit{ensemble of complementary metrics} that capture its different dimensions. This is particularly important in music, where the complexity and interdependence of musical properties make the definition of meaningful composite multi-objective criteria challenging. These metrics enable the evaluation of creativity in AI-generated music and support the definition of new training objectives for generative models, encouraging models to learn what it means to be musically creative in addition to optimizing for the quality of reconstructed artifacts.

As a final note, it must be emphasized that the contribution of this work is conceptual: it proposes and motivates a set of measures, but it does not report an empirical validation of them against human judgments, which is left to future work.

\section{Related Work}

\subsection{Evaluation of AI-Generated Music} The evaluation of AI-generated music takes a plurality of forms. Some approaches focus on assessing audio \textit{quality} using standardized perceptual protocols such as MUSHRA (ITU-R BS.1534) and BS.1116 \citep{itur_2015_bs1534,itu_2015_bs1116}; others examine conditioning through statistical comparisons to training data or stylistic targets, or through specific components such as \textit{pitch} and \textit{rhythm} \citep{lerch_2025_survey,vinay_2022_metrics}. More abstract dimensions such as \textit{originality} and \textit{plagiarism} have been explored via the Originality Report of \citet{yin_2021_good}, which quantifies how much of a generated passage is reproduced from the training corpus, and via the detection of replication of training data in generated audio \citep{batlleroca_2024_replication}.
\textit{Aesthetic} assessment is typically conducted through human preference ratings, with criteria such as naturalness, musicality, and structure evaluated on Likert-type scales \citep{likert_1932_attitudes}, though learned aesthetic predictors trained on human annotations have recently emerged \citep{tjandra_2025_audiobox,yao_2025_songeval}. That the problem remains open is evidenced by its adoption as a shared challenge task
\citep{icassp_2026_aesthetics}.

A recurring finding across this literature is that widely used objective metrics align only weakly with human perception. \citet{groetschla_2025_benchmarking} report that Fréchet Audio Distance correlates with listener preference only for certain embedding backbones, and that metric rankings can diverge substantially from human rankings for individual systems. More recent work notes that the only per-sample metric achieving high correlation with human perception was closed-source, and proposes an open alternative \citep{muqeval_2026}. This is precisely the gap that a principled decomposition of the quantity being measured is meant to address.

\subsection{Subjective Assessment} The first and most common approach is subjective assessment, relying on personal judgment and fuzzy notions of aesthetics \citep{xiong_2023_a}. Broadly speaking, it is possible to discriminate between two types: Turing-style tests, which ask listeners to identify whether a piece was generated by a human or a machine \citep{sturm_2017_models}, and preference tests, where listeners rate outputs along criteria such as musicality, coherence, or naturalness, typically using Likert-type scales \citep{chu_2022_empirical}. While Turing tests are simple and theoretically unbiased, they risk conflating imitation with quality and can be sensitive to stimulus choice and listener expertise. Preference tests, though more nuanced, lack methodological standardization and suffer from overlapping criteria \citep{lerch_2025_survey}. Examples of subjective metrics include Mean Opinion Score (MOS), where participants rate individual audio samples on a scale from 1 to 5 against criteria such as musicality or quality, and Head-to-Head (HTH) comparison, which presents two audio clips side by side and asks participants to choose the better one against a specified criterion \citep{groetschla_2025_benchmarking}. \citet{groetschla_2025_benchmarking} adopted the head-to-head format for their own large-scale benchmarking study, which involved over 15,000 human pairwise evaluations of 6,000 AI-generated music clips across 12 models. Subjective assessments of creativity are human-based and imperfect, yet they remain the most accessible route for researchers and developers.

\subsection{Objective Assessment} The second approach uses objective metrics to analyze features such as audio quality or musical structure, but often neglects qualitative aspects of music \citep{lerch_2025_survey}. To measure audio quality, it is common to use the signal-to-noise ratio, or more sophisticated perceptually motivated methods such as PEAQ \citep{itu_2001_bs1387} and ViSQOL \citep{hines_2015_visqol}. A further family of objective measures draws on music information retrieval and computational musicology, extracting interpretable descriptors of pitch, rhythm, timbre, and structure from the generated signal. Such descriptors have long underpinned tasks like genre classification \citep{tzanetakis_2002_musical} and performance analysis \citep{lerch_2019_music}, and they transfer naturally to generative evaluation, though the field has repeatedly noted the difficulty of establishing shared ground truth and evaluation protocols \citep{downie_2004_scientific}. In general, four strategies are available: comparing distributions between the generated artifact and the training data; pairing the generated file with a reference to examine reconstruction; measuring properties of the generated music without any reference; and using classification algorithms to evaluate the artifact \citep{lerch_2025_survey}. Adopting an objective approach to evaluating AI-generated musical creativity is not only possible but necessary \citep{oneill_2017_limitations}.

\subsection{Rubric-Based Evaluation} A pragmatic middle ground between purely subjective and purely objective assessment has emerged in the training of large language models, where subjective qualities are operationalized by having domain experts co-design \textit{rubrics}: collections of fine-grained criteria that are checked, often in binary fashion, and aggregated into a scalar reward signal used during post-training \citep{ouyang_2022_training,mu_2024_rule,gunjal_2026_rubrics}. In this paradigm, the individual criteria are typically adjudicated by an LLM grader prompted with the output and the criterion \citep{mu_2024_rule,gunjal_2026_rubrics}, and such rubric-based rewards have been shown to extend reinforcement learning beyond
domains with a single verifiable answer \citep{gunjal_2026_rubrics}. This practice is increasingly a service in its own right, supplied at scale by commercial data-annotation providers such as Scale AI and Mercor, both of which have published expert-authored, rubric-based benchmarks graded by LLM judges \citep{akyurek_2026_prbench,vidgen_2025_apex}. The musical analogue would be a battery of expert-authored questions---for instance, ``does the melody contain a note outside the expected scale?''---aggregated into a composition score, echoing long-standing rubric-based assessment in music pedagogy \citep{hickey_1999_assessment}. This approach faces three limitations that motivate the one proposed here. First, it does not scale cleanly to audio: it is unclear whether each criterion should be adjudicated by an audio-capable model prompted with the clip and the question, or computed by a bespoke signal-level formula, and neither route is systematic across criteria. Second, expert-authored rubrics are expensive, requiring a large panel of annotators to assemble a dataset of
usable size, and must be relabeled as models and objectives evolve \citep{mu_2024_rule}. Third, there is no guarantee that optimizing such an aggregate systematically increases musical \textit{creativity} rather than surface conformance to a checklist. These shortcomings motivate \textit{verifiable}, reproducible ways to quantify subjective musical qualities, which is the aim of the framework this paper proposes.

\subsection{Measuring Creativity Itself} \textit{Creativity}, often described as an ``ill-defined'' concept \citep{lerch_2025_survey}, lacks robust assessment methods. General frameworks such as the SPECS model of \citet{jordanous_2012_evaluating} encourage evaluation along defined dimensions including skill, novelty, and value, but typically require human involvement and are not turnkey solutions, which has limited their adoption in music evaluation. Other theoretical models, including FACE and IDEA \citep{colton_2011_face}, aim to formalize the assessment of creative output but fall short in empirical applications. The FACE model formalizes a creative act as a non-empty tuple of generative acts, each comprising zero or one instance of eight types such as concept generation (Cg), expression generation (Eg), aesthetic measure generation (Ag), and framing information (Fg); a tuple $\langle$Fg, Ag, Cg, Eg$\rangle$, for instance, represents a creative act in which an expression is produced by a concept, assessed by an aesthetic measure, and further explained by framing information \citep{colton_2011_face}. Complementing this, the IDEA model evaluates the impact of such creative acts on an idealized audience through measures of well-being and cognitive effort \citep{colton_2011_face}.

Closer to this paper's approach are attempts to operationalize Boden's three criteria computationally. \citet{franceschelli_2022_deepcreativity} propose DeepCreativity, a modular measure of value, novelty, and surprise computed without human judgment, demonstrated on the generation of nineteenth-century American poetry. Others have used the triad, or parts of it, to steer generation rather than as an evaluation target: surprise search rewards candidates that deviate from the outcomes the generator itself expects \citep{yannakakis_2016_searching}; a music-motivated composer--audience architecture learns to produce events that violate the expectations of a separately trained audience model \citep{bunescu_2019_learning}; Chef Watson selects among generated recipes using a flavor-pleasantness model for quality and Bayesian surprise for novelty \citep{varshney_2019_big}; and Creative Preference Optimization injects novelty, surprise, diversity, and quality signals into the preference-optimization objective of LLMs \citep{ismayilzada_2025_creative}. None of these efforts adapts value, novelty, and surprise jointly to the specific structure of musical artifacts, which is the gap this paper addresses.

The adequacy of the three-criteria decomposition has itself been questioned. \citet{salamanca_2023_dynamic} asked 43 listeners of varying musical training to judge unfinished pieces produced by a co-creative assistant, and found that value and novelty alone sufficed to predict creative merit, with novelty and surprise rated statistically indistinguishably. Here, this result is taken seriously but read differently. Their stimuli were \textit{proto-artifacts} judged by human raters on a single global scale, a setting in which novelty and surprise are difficult for a listener to separate introspectively. The paper's perspective is that the two come apart under computational measurement of finished artifacts: a piece may diverge sharply from a reference corpus while remaining internally predictable, and conversely may sit well within a familiar style while systematically violating local expectations. Sections~\ref{sec:msurprise} and~\ref{sec:mnovelty} make this distinction explicit, and the decomposition of surprise into probabilistic, hidden-content, and paradigm-shifting varieties identifies phenomena that no novelty measure captures. Whether the two measures are empirically separable in practice is an open question.

Overall, to the best of our knowledge, there is no general framework for quantitatively
evaluating AI-generated musical creativity \citep{lerch_2025_survey}. The proposed framework is intended to conceptually fill that gap: rather than adding another quality or similarity measure, it decomposes creativity itself into an ensemble of complementary components that can be operationalized and computed at the artifact level.

\section{Defining AI Creativity in Music}

In order to define AI creativity in music, a conceptualization of music and creativity must first be provided. It is important to acknowledge that the definitions drawn here can be subject to significant enhancements.

\subsection{AI-Generated Music}

\subsubsection{On Defining Music: An Interdisciplinary View}

Music feels so intuitive that most people do not feel the need to define it. Yet, the concept of music is widely debated among scholars and defined differently across disciplines \citep{hallam_2022_the}. A thoughtful approach to understanding music is that of \citet{davies_2012_on}, who distinguishes four different ways of defining it: \textit{functionally}, \textit{operationally}, \textit{historically}, and \textit{structurally}.

Defining music functionally consists of looking at the various functions music has in its different forms and genres in relation to humans. For instance, sad songs and love songs aim to evoke or regulate specific human emotions like sorrow or longing \citep{davies_2012_on}. Muzak, often referred to as elevator music, is designed to encourage purchases by creating a specific ambiance \citep{anderson_2015_neomuzak}. Additionally, music plays a crucial role in enhancing productivity and focus, with instrumental or ambient tracks being popular for studying \citep{kotsopoulou_2010_the}, while upbeat music is commonly used to boost physical exercise, and nightclub music to foster social interactions. Of course, music has many more functions \citep{schafer_2013_psychological}.

Operationally, music can be defined by its ability to activate specific neural circuits in the brain, focusing on measurable effects rather than its essence \citep{davies_2012_on}. Studies in psychology and neuroscience \citep{levitin_2006_this, patel_2008_music} have shown that music stimulates areas linked to emotion, memory, and reward. However, defining music solely by its neural effects is challenging since it relies on neuroimaging and other brain activity measurement techniques, which, while increasingly sophisticated, remain limited in their ability to fully capture the complexity of neural processing \citep{chan_2022_the}.

Structurally, music is most often defined in terms of its elements and the relationships that arise from their combination. It aims to identify ‘atomic’ musical units, such as pitched tones, and describe how these combine into higher-order structures like phrases, melodies, and movements \citep{davies_2012_on}. This approach is reminiscent of linguistic generative grammars, where phonemes combine into words, words into phrases, and phrases into sentences, governed by syntactic and semantic rules \citep{chomsky_1957_syntactic,lerdahl_1983_generative}. Related to this are music universals, features shared across musical traditions that help distinguish music from non-music. A key example is octave equivalence, the perceptual phenomenon where notes an octave apart (e.g., C3 and C4) are heard as versions of the same pitch, despite their frequency doubling \citep{brown_2013_universals}. This principle reflects fundamental aspects of human auditory perception and underpins scale construction in diverse cultures. However, not all music is tonal or hierarchically structured (e.g., the use of untuned instruments), and structural approaches often overlook the cultural and social practices that give music its significance. Music is a culturally embedded human practice shaped by historical traditions, sociocultural norms, and psychological responses \citep{davies2007essential, davies_2012_on}.

Historically, music is defined by its evolution within sociocultural contexts, shaped by changes in composition, technology, and societal roles. Western music, for example, progressed from monodic singing to polyphony and complex tonal systems, reflecting shifts in musical language \citep{davies_2012_on}. Yet, again, defining music through a historical lens does not fully capture its impact on human well-being or explain why certain compositions, like Mozart’s, have endured despite significant societal transformations. Music, unlike many other human activities, appears to function as a universal language that has a certain ‘power’ \citep{hallam_2022_the}.

In sum, music is highly universal and transcends disciplines and domains of application. From neurosciences and psychoacoustics (operational), psychology (functional), music theory and mathematics (structural), or sociology and history (socio-historical), a different perspective is offered \citep{schiavio_2021_editorial}.

\subsubsection{Introduction to Artificially Intelligent Music Generation}

\begin{figure}[t]
    \centering
    \includegraphics[width=0.75\textwidth]{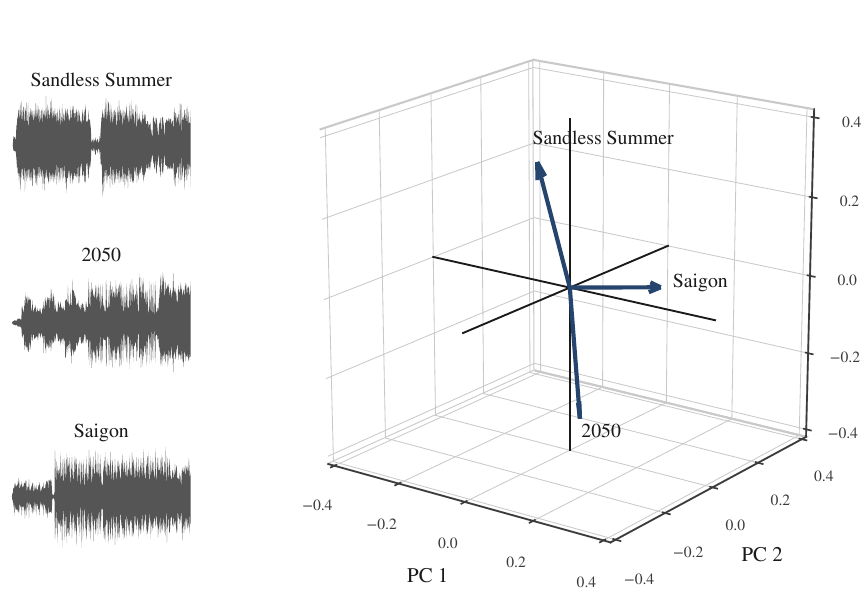}
    \caption{3D embedding vector representation of three commercially released music pieces \citep{hindy2050,hindy_saigon,hindy_sandless} from CLAP \citep{elizalde_2023_clap}, after dimensionality reduction.}
    \label{fig:embedd}
\end{figure}

Music generation typically follows two approaches: symbolic and audio-based \citep{xiong_2023_a}. In symbolic generation, music is represented as discrete sequences of musical events, such as MIDI files or music scores, capturing information like pitch, duration, and velocity. This approach is generally more interpretable and computationally efficient. Notable models include the Music Transformer \citep{huang_2019_music} and MidiNet \citep{yang_2017_midinet}. In contrast, audio generation models work directly with sound signals. Early approaches used hand-crafted features such as spectrograms and Mel-frequency cepstral coefficients (MFCCs) to represent musical elements like pitch, rhythm, and timbre \citep{tzanetakis_2002_musical}. More recently, advances in representation learning have enabled models to work directly with raw audio, using neural audio codecs like Encodec \citep{defossez_2023_high} and self-supervised representation models like MERT \citep{li_2024_mert} to learn compressed, perceptually meaningful embeddings (see an example in Figure~\ref{fig:embedd}). Systems such as MusicLM \citep{agostinelli_2023_musiclm}, MusicGen \citep{copet_2023_simple}, and Stable Audio Open \citep{evans_2025_stable} use these embeddings to generate high-quality audio conditioned on text or symbolic prompts.

Music generation has also evolved in its use of input modalities \citep{li_2026_survey}. Early models focused exclusively on musical data, performing single-modal generation tasks such as symbolic-to-symbolic generation (e.g., diffusion over symbolic representations \citep{mittal_2021_symbolic}) or audio-to-audio transformation (e.g., VampNet \citep{garcia_2023_vampnet}). Later, models began incorporating external modalities like text, images, and video to guide the music generation process, leading to cross-modal generation. Examples include text-to-music models such as MusicLM \citep{agostinelli_2023_musiclm}, image-to-music models like MeLFusion \citep{chowdhury_2024_melfusion}, and video-to-music systems such as V2Meow \citep{su_2024_v2meow}. 

Today, the frontier is multi-modal generation, where multiple inputs, including text, images, and video, are combined to create richer and more controllable musical outputs, as seen in systems like Seed-Music \citep{bai_2024_seedmusic} and MuMu-LLaMA \citep{liu_2026_mumullama}. These systems typically encode each input into a latent space using models like VQ-VAEs (audio), Transformers (text), and ViTs (images), fuse them via cross-attention or joint embeddings, and condition a generation model (GANs, diffusion, or Transformers) to produce music \citep{li_2026_survey}.

\subsection{AI-Generated Musical Creativity}

\subsubsection{On Defining Creativity: From Humans to Generative Artificial Intelligence}

So far, the computational approach to defining creativity has avoided investigating the semantic meaning of the word. For instance, \citet{colton_2012_computational} define Computational Creativity (CC) as ‘the philosophy, science and engineering of computational systems which, by taking on particular responsibilities, exhibit behaviors that unbiased observers would deem to be creative.’ This issue is at the core of the creative challenge of machines for two main reasons. 

First, Colton and Wiggins’ definition of CC necessarily invokes human observers; hence, it is uniquely subjective. There is no explicit formulation of some objective characteristics of a creative artifact. Second, there is no domain-specific framing of creativity, despite such a phenomenon being highly activity-dependent (e.g., being creative in composing music is not the same as being creative in cooking). This challenge faced in the field of CC derives from an active debate on defining creativity. For example, while \citet{hennessey1988conditions} emphasize that creativity involves originality with a purpose, a sort of problem-solving ability, the Torrance Tests of Creative Thinking \citep{torrance_1966_tests} operationalize creativity as divergent thinking, measuring how far individuals can generate new ideas from existing information.

\subsubsection{Creativity: Value, Novelty, and Surprise} \label{sec:boden}

In the context of AI systems, Margaret Boden, a pioneer in the field of CC, proposes an output-focused framework to describe creativity, which is particularly relevant to the goal of measuring the creativity of ‘black box’ models, whose inner processes often escape our understanding. Boden breaks the concept of creativity into three components: \textit{value}, \textit{novelty}, and \textit{surprise} \citep{boden_2004_the}. 

At a high level, value refers to the extent to which an idea is considered useful and appropriate within a given context. Novelty refers to how different an idea instantiated by a given object is, either relative to all ideas generated in the past (historical novelty) or relative to all ideas generated by the object itself (psychological novelty). Finally, surprise captures the extent to which an idea deviates from expectations based on prior knowledge or experience.

Additionally, Boden distinguishes between three forms of creativity: \textit{combinational} (C-C), \textit{exploratory} (E-C), and \textit{transformational} (T-C). Combinational creativity involves generating ideas by combining familiar concepts in new ways. Exploratory creativity refers to exploring a structured conceptual space (set of rules and constraints) to produce new but valid ideas within that space. Transformational creativity comes from altering the underlying rules or constraints themselves, thereby redefining the conceptual space. T-C can also be seen as meta-level E-C, as it involves exploring not just ideas within a system, but systems within a larger set of systems.

In the literature, some systems claim to achieve transformational creativity but fall short. For example, \citet{hung_2013_conceptual} Conceptual Recombination (CR) model recombines user inputs using fixed rules and preserved structures, allowing only surface-level changes. As a result, it stays within a constrained conceptual space, aligning more with combinational than transformational creativity in the strict Bodenian sense \citep{boden_2004_the}.

\subsubsection{The Curse of Subjectivity in Creativity}

Assessing the creativity of a piece of music is not a matter of simply classifying it as ‘creative’ or ‘non-creative’. Instead, creativity exists on a continuous spectrum, where each listener assigns a different degree of creativity to the same piece based on their individual perception. Because of such an observer-dependent nature, the same piece of music may be judged as highly creative by one person and as uninspired by another, depending on their background, expertise, and familiarity with the style.

Today, tools like Suno AI \citep{suno} may seem highly creative to non-musicians, but music experts apply stricter standards, making it harder for AI systems to ‘fool’ them. At the same time, expert judgment is constrained by familiarity with specific musical styles. Unlike most humans, AI models can generate music across a wide range of genres. Consider a professional jazz musician evaluating an AI model: because they have spent years listening to, performing, composing, and improvising jazz, they are likely to find the model’s jazz compositions unoriginal. However, if the same jazz musician were to evaluate AI-generated hip-hop music, a genre they have little experience engaging with, they might be much more impressed and perceive the output as highly creative. Therefore, creativity is subjective, observer-dependent, and task-dependent.

\subsubsection{Definition of AI Musical Creativity} \label{sec:mcreativity-definition}

Akin to creativity in the general Bodenian sense (see Section~\ref{sec:boden}), Generative Artificial Intelligence's Musical Creativity (\textbf{M-Creativity}) can be understood as the interplay between three main components: 

\begin{itemize}
    \item \textbf{M-value}, which quantifies the musical worth of a composition by a listener. In other words, M-value corresponds to how valuable a music artifact is perceived by the listener. 

    \item \textbf{M-surprise}, which captures the extent to which a listener’s musical expectations are violated or challenged by a piece of music.

    \item \textbf{M-novelty}, which refers to the degree of newness, often synonym for deviation of the generated music from the training corpus or from established norms. 
\end{itemize}

In theory, when one of these components takes a high value relative to the norm (as computed from large, representative music datasets), it is believed to contribute positively to the M-Creativity score. Assuming all three components are non-negative, it is possible to distinguish the cases summarized in Table~\ref{tab:mcreativity_cases}. In the remainder of the paper, the components are treated in the order M-value (Section~\ref{sec:mvalue}), M-surprise (Section~\ref{sec:msurprise}) and M-novelty (Section~\ref{sec:mnovelty}), because the treatment of novelty builds on the distinction from probabilistic surprise established in Section~\ref{sec:msurprise}.

\begingroup
\small
\renewcommand{\arraystretch}{1.35}
\begin{longtable}{@{}L{\dimexpr 0.37\linewidth-\tabcolsep\relax} L{\dimexpr 0.63\linewidth-\tabcolsep\relax}@{}}
  \caption{Combinatorial Effects of M-value, M-novelty, and M-surprise on M-Creativity}
  \label{tab:mcreativity_cases}
  \\
  \toprule
  \textbf{Component Configuration} & \textbf{Interpretation and Effect on M-Creativity} \\
  \midrule
  \endfirsthead

  \multicolumn{2}{c}{\textbf{\tablename\ \thetable} — continued from previous page} \\
  \toprule
  \textbf{Component Configuration} & \textbf{Interpretation and Effect on M-Creativity} \\
  \midrule
  \endhead

  \midrule \multicolumn{2}{r}{\textit{Continued on next page}} \\
  \endfoot

  \bottomrule
  \endlastfoot

  \textbf{All components above norm} \newline
  ($\mathrm{M\text{-}value}>\mu_V$, $\mathrm{M\text{-}novelty}>\mu_N$, $\mathrm{M\text{-}surprise}>\mu_S$) &
  Represents ideal creativity. The musical output is valuable, original, and surprising, typically perceived as highly creative and impactful. \\

  \textbf{One component far above norm} \newline
  (e.g., $\mathrm{M\text{-}value}\gg\mu_V$, while others $\approx \mu$) &
  A very strong component may compensate for average or weak others. Creativity can still be perceived as high, especially if the dominant trait is M-value. \\

  \textbf{One component below norm} \newline
  (e.g., $\mathrm{M\text{-}value}<\mu_V$, others $>\mu$) &
  Creativity may be reduced despite novelty and surprise. The work might be interesting but aesthetically weak, dissonant, or incoherent. \\

  \textbf{Two components below norm} \newline
  (e.g., $\mathrm{M\text{-}novelty}<\mu_N$, $\mathrm{M\text{-}surprise}<\mu_S$) &
  If only M-value is above norm, output may be technically sound but uninspired. Creativity is typically below average. \\

  \textbf{All components below norm} \newline
  ($\mathrm{M\text{-}value}<\mu_V$, $\mathrm{M\text{-}novelty}<\mu_N$, $\mathrm{M\text{-}surprise}<\mu_S$) &
  Represents a failure of creativity. The result is likely predictable, unoriginal, and musically unengaging. \\

  \textbf{Balanced trade-off or synergy} \newline
  (e.g., $\mathrm{M\text{-}novelty}>\mu_N$, $\mathrm{M\text{-}surprise}>\mu_S$, $\mathrm{M\text{-}value}\approx\mu_V$) &
  Moderate or near-norm value in one component can be offset by strong performance in others. Creative synergy emerges when two components jointly elevate the perception of novelty and impact. \\

\end{longtable}
\endgroup

\section{A Framework for Quantifying AI-generated Musical Creativity} \label{sec:mcreativity}

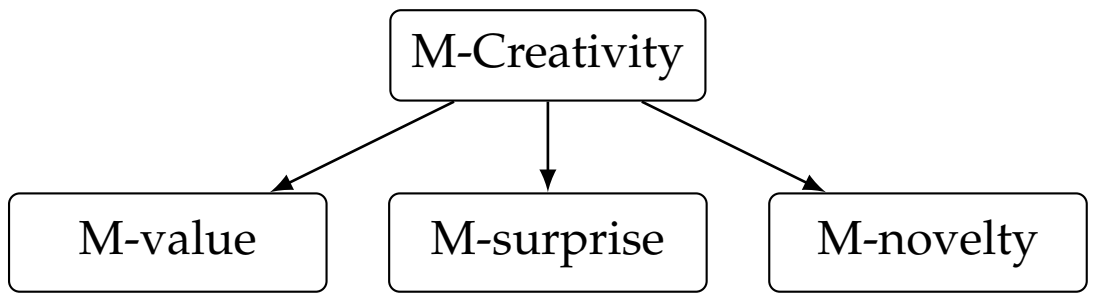
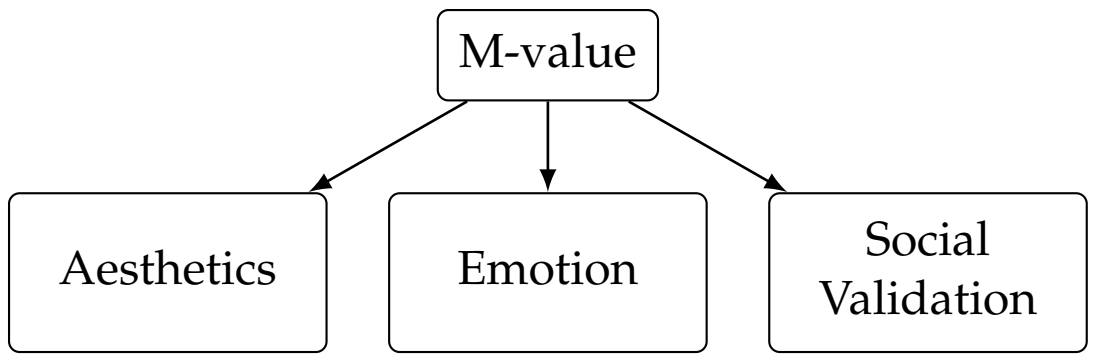
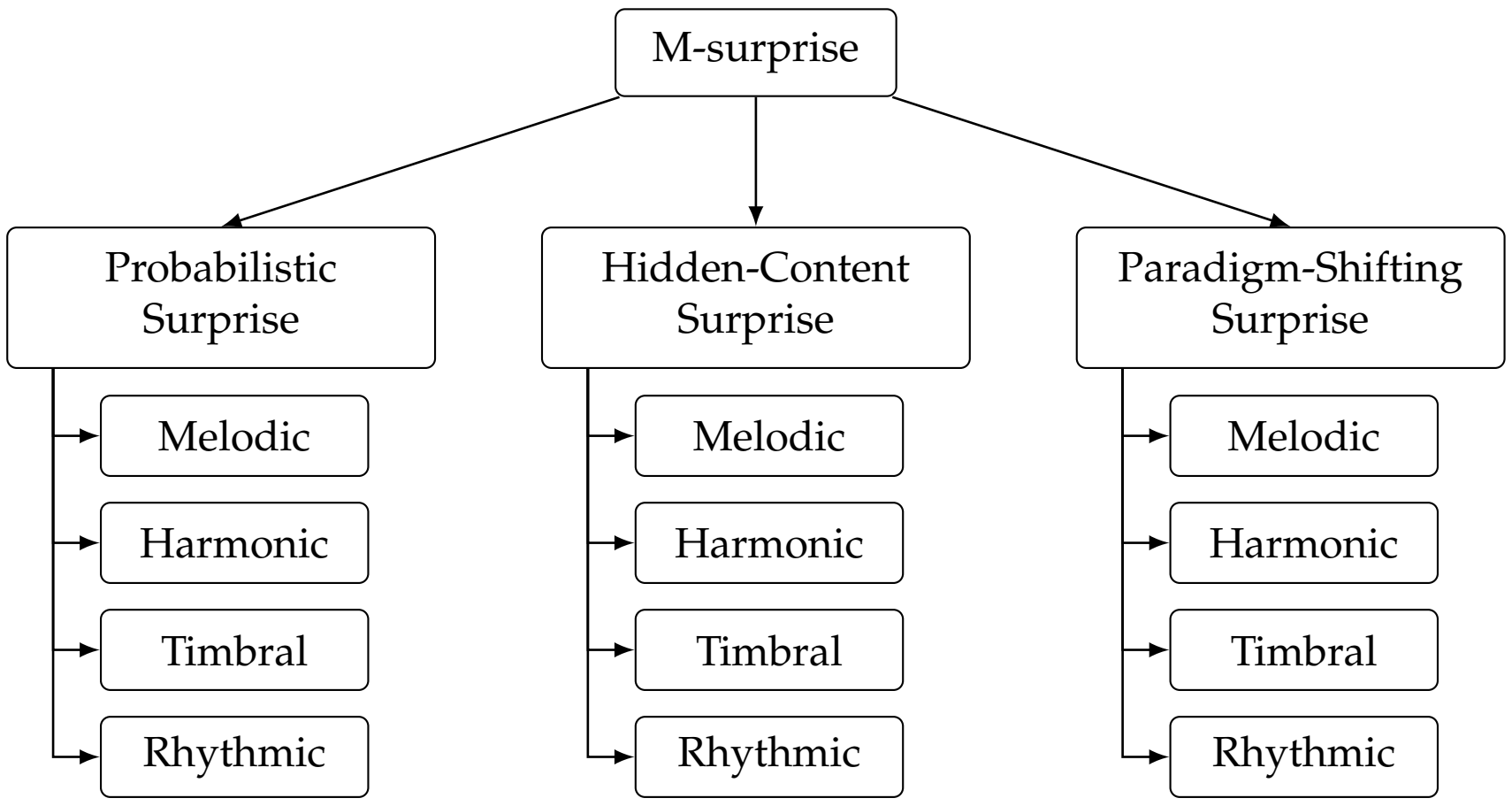
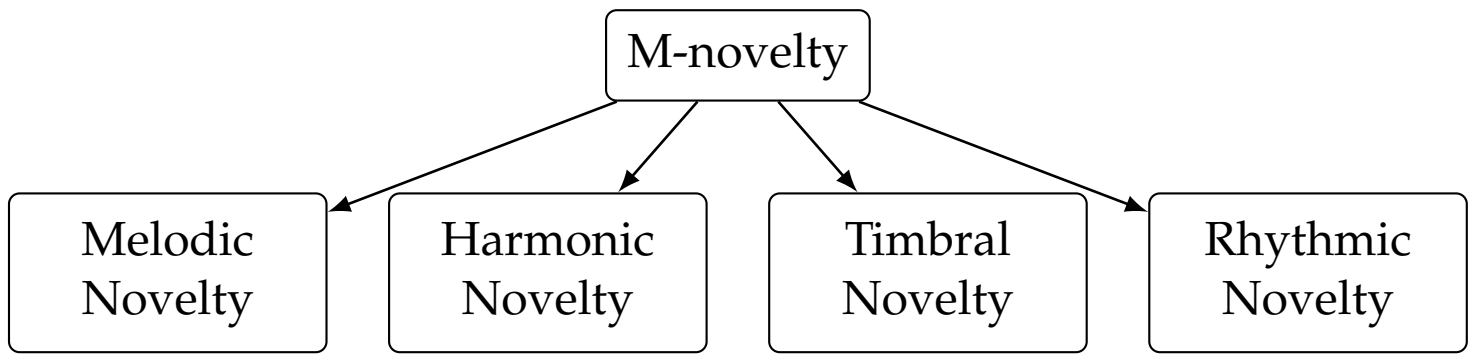
\begin{figure}[!htbp]
\centering

\begin{subfigure}{\textwidth}
  \centering
  \begin{forest} mc/base, where level=1{mc/one}{}
    [M-Creativity
      [M-value]
      [M-surprise]
      [M-novelty]
    ]
  \end{forest}
  \caption{M-Creativity (top-level components)}
\end{subfigure}

\vspace{0.8em}

\begin{subfigure}{\textwidth}
  \centering
  \begin{forest} mc/base, where level=1{mc/two}{}
    [M-value
      [Aesthetics]
      [Emotion]
      [Social\\Validation]
    ]
  \end{forest}
  \caption{M-value}
\end{subfigure}

\vspace{0.8em}

\begin{subfigure}{\textwidth}
  \centering
  \begin{tikzpicture}[
      mcbox/.style={draw, rounded corners=2pt, align=center, font=\small,
                    inner xsep=4pt, inner ysep=3pt},
      mcarrow/.style={-latex, line width=0.5pt}]
    \node[mcbox, minimum width=21mm, minimum height=6.5mm] (ms) at (0,0) {M-surprise};
    \foreach \nm/\lbl/\x in {p/{Probabilistic\\Surprise}/-40mm,
                             hc/{Hidden-Content\\Surprise}/0mm,
                             ps/{Paradigm-Shifting\\Surprise}/40mm}{
      \node[mcbox, anchor=north, minimum width=32mm, minimum height=10.5mm]
        (\nm) at (\x,-13mm) {\lbl};
      \draw[mcarrow] (ms) -- (\nm.north);
      \foreach \dim [count=\i] in {Melodic, Harmonic, Timbral, Rhythmic}{
        \node[mcbox, anchor=west, minimum width=20mm, minimum height=6mm]
          (\nm\i) at ([xshift=7mm, yshift={-(\i*8mm-3mm)}]\nm.south west) {\dim};
        \draw[mcarrow] ([xshift=3.5mm]\nm.south west) |- (\nm\i.west);
      }
    }
  \end{tikzpicture}
  \caption{M-surprise}
\end{subfigure}

\vspace{0.8em}

\begin{subfigure}{\textwidth}
  \centering
  \begin{forest} mc/base, where level=1{mc/two}{}
    [M-novelty
      [Melodic\\Novelty]
      [Harmonic\\Novelty]
      [Timbral\\Novelty]
      [Rhythmic\\Novelty]
    ]
  \end{forest}
  \caption{M-novelty}
\end{subfigure}

\caption{Components of the proposed M-Creativity framework.}
\label{fig:framework}
\end{figure}

The proposed evaluation framework is based on Boden's creativity theory (See Section~\ref{sec:boden} for more details on her framework) \citep{boden_2004_the,boden_2009_computer}, and extends it to the music domain, introducing Musical Value (M-value), Surprise (M-surprise), and Novelty (M-novelty). The framework is designed to be modular and extensible, allowing for the addition of new metrics as research progresses. Each component of M-Creativity is broken down into sub-components that can be measured using specific quantitative metrics. The framework is visualized in Figure~\ref{fig:framework}, with each main component (M-value, M-surprise, M-novelty) further divided into measurable sub-components.

\subsection{Tribalism in Prior Musical Experience (PME)}

Before introducing the framework, a key assumption should be made: that musical appreciation is inherently shaped by cultural variation in music perception \citep{mcdermott_2016_indifference}, which is referred to here as Tribalism in prior musical experience. This refers to the idea that listeners within a shared sociocultural group develop similar musical expectations and evaluative criteria, which in turn shape their perception of creativity \citep{frith_1996_performing}. The perceived level of M-Creativity of a musical artifact depends heavily on the listener’s prior musical experience. \citet{vuust_2022_music} describe this personal knowledge accumulation and expectation framework as the ‘brain model’, which guides how listeners predict and respond to musical events. To generalize M-Creativity across audiences, this paper assumes that members of a shared sociocultural and demographic group possess similar brain models. This shared knowledge is approximated through the construct of \textit{Prior Musical Experience} (PME). PME is best understood as an umbrella concept that integrates multiple components, such as:

\begin{itemize}
    \item \textit{Prior Music Listening (PML)}: the body of music a listener has encountered passively or actively throughout life.
    \item \textit{Prior Music Practice (PMP)}: the listener’s experience playing, composing, or formally studying music.
    \item \textit{Other Factors}: including cultural context, physiological predispositions, etc.
\end{itemize}

At present, PML is used as a proxy for PME, conscious that the current approach does not explicitly account for individual differences in PMP or other variables. To estimate the PML (and by extension PME) of a group $L$ of listeners born in a given year $Y$, two sources of exposure are combined:
\begin{itemize}
\item \textit{Temporal Exposure ($I$):} The set $I$ of all music released during a historical window shaped by the listener’s age $A$, birth year $Y$, and a look-back horizon $\delta$ (in years):
\begin{equation}
I = \{\, x \;:\; \mathrm{release}(x) \in [Y - \delta,\ Y + A] \,\}
\end{equation}
This captures music likely encountered via radio, streaming, or cultural transmission. The horizon $\delta$ is a parameter that should be set by the researcher according to the type of music under study: genres with long-lived canonical repertoires warrant a larger $\delta$, whereas fast-moving contemporary styles may call for a smaller one. As an illustrative default, this paper takes $\delta = 50$ years, reflecting the intergenerational transmission of music within a household and community---a listener is routinely exposed not only to the music of their own era but also to that of their parents' and grandparents' generations. For example, with $\delta = 50$, a listener born in 2000 and aged 25 is assumed to be exposed to music from 1950 to 2025.

\item \textit{Timeless Culturally Significant Music (TCSM)}
Seminal compositions with enduring cultural relevance. These vary by region (e.g., for Western European listeners, pieces like Vivaldi's \textit{The Four Seasons} \citep{vivaldi1725score}).
Thus, the group-level prior music listening set is defined as:
\begin{equation}
\text{PML} = I \cup \text{TCSM}
\end{equation}
In practice, the $I$ set can be constructed using large historical databases such as the Million Song Dataset \citep{bertinmahieux_2011_million} or Discogs \citep{discogs_2010_discogs}, filtered by release date. The TCSM set can be derived from curated playlists like Spotify’s \textit{Essentials of Classical Music} \citep{spotify_2025_essentials}, offering a reasonable estimate of widely shared musical exposure.
\end{itemize}

\section{M-value} \label{sec:mvalue}

As noted by \citet{boden_2004_the}, the concept of ‘value’ encompasses an unbounded and highly variable set of meanings; it is a one-to-many mapping whose sense shifts over time and resists a single definition. Rather than an intrinsic property of an object or idea, value is always relative to an observer and a context: it is more precise to say that something is valuable \textit{with respect to someone, or something} than that it is valuable in itself.

\subsection{Aesthetics as a Conceptual Basis for M-value}

Margaret Boden's definition of value \citep{boden_2004_the} is rooted in the concept of \textit{aesthetics}, a notion that has long been considered notoriously difficult to define: 

\begin{quote}
    What’s difficult is to state our aesthetic values clearly enough to enable the program itself to make the evaluation at each generation. \citep[p.~25]{boden_2009_computer}
\end{quote}

\citet{himonides_2009_mapping} proposes that the perception of musical beauty, and thus its value with respect to the listener, is shaped by seven interrelated components (see Table~\ref{tab:aesthetic_components}). While originally concerning the singing voice, they can be generalized to music at large. These seven components interact to shape what listeners find aesthetically valuable in music. However, there is no universal formula for combining them into a single M-value score. Each listener may assign different weights to these components, and these weights can change over time, across cultures, and depending on context \citep{deschenes1989creativity}. Moreover, genres evolve, norms shift, and cultural expectations transform.


\begingroup
  \small
  \renewcommand{\arraystretch}{1.35} 
  \begin{longtable}{@{}L{\dimexpr 0.33\linewidth-\tabcolsep\relax} L{\dimexpr 0.67\linewidth-\tabcolsep\relax}@{}}
  \caption{Components of Aesthetic Perception in Music, adapted from \citet{himonides_2005_building,himonides_2009_mapping}}
  \label{tab:aesthetic_components}
  \\
  \toprule
  \textbf{Component} & \textbf{Description} \\
  \midrule
  \endfirsthead
  
  \multicolumn{2}{c}{\textbf{\tablename\ \thetable} - continued from previous page} \\
  \toprule
  \textbf{Component} & \textbf{Description} \\
  \midrule
  \endhead
  
  \midrule \multicolumn{2}{r}{\textit{Continued on next page}} \\
  \endfoot
  
  \bottomrule
  \endlastfoot
  
  \textbf{Musical Structure} \newline \citep{ockelford_2009_zygonic, scherer2001emotional} &
  Formal elements such as motifs, harmony, phrasing, and chord progressions shape emotional arcs and guide listener expectations. The I–V–vi–IV progression is a widely used and emotionally impactful example. \\
  
  \textbf{Neuropsychobiological Processing} \newline \citep{peretz2004singing} &
  Music perception is rooted in how the brain processes sound. Neural circuits involved in pitch and rhythm overlap with those used in language. Responses vary based on experience, emotion, and training \citep{sarasso_2019_aesthetic}. \\
  
  \textbf{Human Musical Development} \newline \citep{welch2005singing, malloch1999mothers} &
  Musical understanding arises from biological predisposition and cultural exposure. Even without training, listeners acquire intuitive knowledge of musical norms. \\
  
  \textbf{Sociocultural Context} \newline \citep{himonides_2005_building, durrant_2017_choral} &
  Cultural backgrounds shape what is valued in music. Different genres emphasize different features (e.g., timbre in jazz, rhythm in rap), forming genre-specific aesthetic expectations. \\
  
  \textbf{Listener Characteristics} \newline \citep{sloboda2001emotions} &
  Personal factors such as emotion, personality, and memory shape perception. The same piece may be valued differently depending on the listener’s context. \\
  
  \textbf{Production of Acoustic Signals} \newline \citep{sundberg_1987_science, juslin2001communicating, gabrielsson_2003_music} &
  Vocal performance is shaped by control of breath, vocal fold tension, and resonance, which affect tone and expressiveness. \\
  
  \textbf{Effects of Acoustics} \newline \citep{howard_2017_acoustics} &
  Performance spaces and recording setups (e.g., reverb, mic placement) influence how music is perceived in terms of clarity and impact. \\
  
  \end{longtable}
  \endgroup

When considering musical aesthetic as a proxy for M-value, it is also possible to use certain objective metrics previously proposed in the literature, such as computational aesthetics measures grounded in information and complexity theory, geometric measures that analyze statistical distributions of musical elements, or biologically inspired measures that explore social, genetic, and evolutionary algorithms \citep{kalonaris_2018_computational}. Additionally, Zipf's law has been found to relate to musical aesthetic preferences, with research suggesting that music whose element distributions are closer to Zipfian is judged more aesthetically pleasing \citep{manaris_2005_zipf}. Zipfian conformity is assessed by ranking the musical elements (e.g., pitches, durations, intervals) by frequency of occurrence and fitting the rank--frequency relation
\begin{equation}
f(n) \propto n^{-\alpha},
\end{equation}
\noindent where $n$ is the rank of the element (the most frequent element has rank $1$) and $f(n)$ its frequency. In practice, $\alpha$ is estimated as the negative slope of a linear regression of $\log f(n)$ on $\log n$; a slope close to $-1$ (and a high coefficient of determination) indicates a Zipfian distribution \citep{manaris_2005_zipf}.

\subsection{Emotion as Conceptual Basis for M-value}

\begin{figure}[ht!]
  \centering
  \begin{tikzpicture}[>=stealth, line width=0.8pt, scale=0.4]

      \draw[dashed, gray] (0,0) circle (5.4cm);
      
      \draw[->] (0,-5.5) -- (0,5.5) node[above]{High Arousal};
      \draw[->] (-5.5,0) -- (5.5,0) node[right]{High Valence};
      \node[below] at (0,-5.5) {Low Arousal};
      \node[left] at (-5.5,0) {Low Valence};
      
      \node at (-3,3) {Tense};
      \node at (3,3) {Excited};
      \node at (-3,-3) {Depressed};
      \node at (3,-3) {Relaxed};
      
  \end{tikzpicture}
  \caption{Valence–Arousal Model of Emotion, adapted from \citet{russell_1980_a}.}
  \label{fig:valence_arousal}
\end{figure}
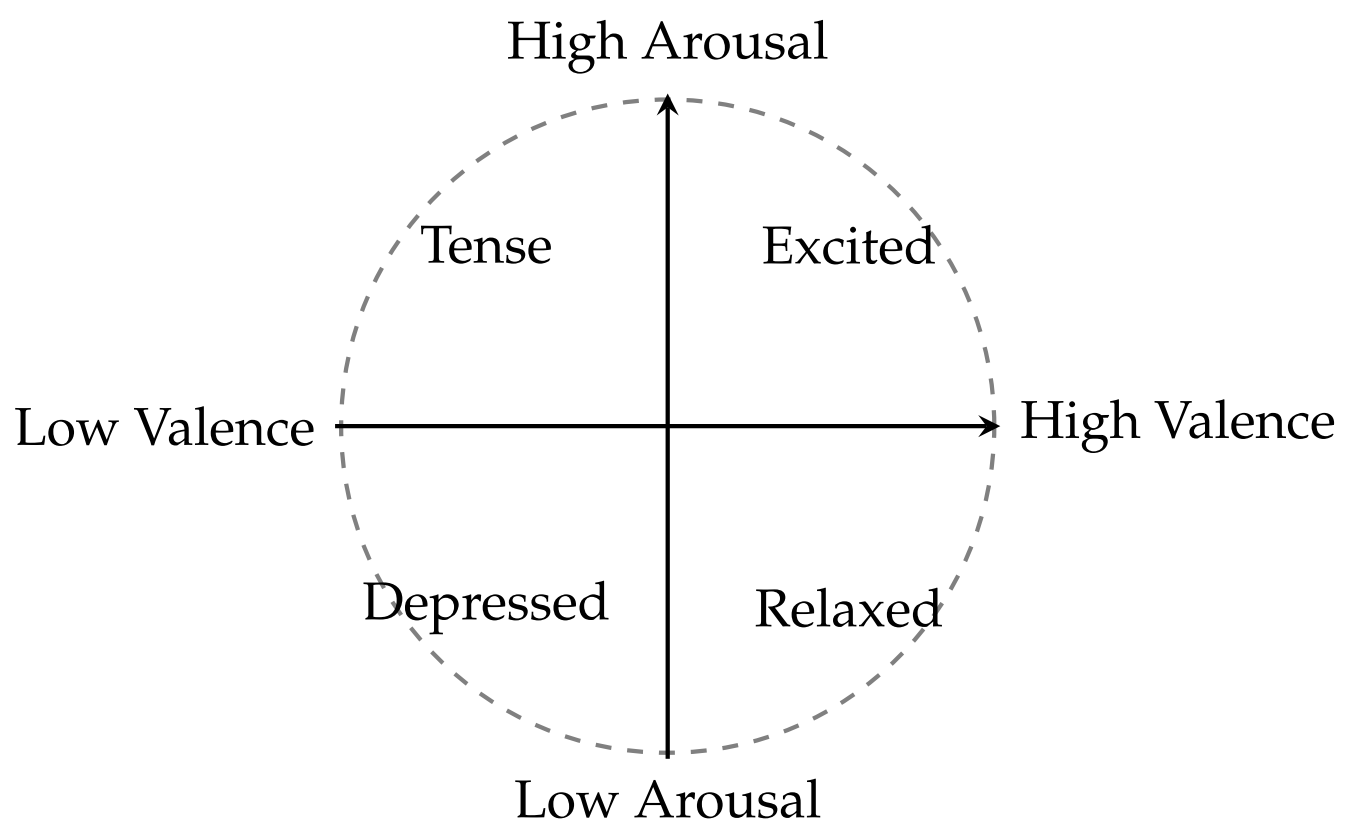

The listener values music that triggers certain emotions \citep{gurgen_2016_social}. One approach to quantifying musical value for a listener is to analyze the level of emotional content elicited by music. In this regard, Wundt’s model \citep{berlyne1970novelty, wundt1904principles} conceptualizes emotions along three primary dimensions (pleasure–displeasure, arousal–relaxation, and strain–relief), offering an early framework for understanding emotional responses. However, Wundt’s model \citep{berlyne1970novelty, wundt1904principles} is largely introspective and lacks the applicability required by modern affective research \citep{russell_2003_core}.
A more empirically grounded alternative is the Valence–Arousal model \citep{russell_1980_a}, which reduces emotional experience to two neurophysiologically supported axes: valence (positive–negative) and arousal (high–low) (See Figure~\ref{fig:valence_arousal}). This dimensional model has been validated through a broad range of psychological and neuroscience studies \citep{larsen1992circumplex, russell1999core}, and aligns closely with physiological and neurological markers of emotion such as EEG asymmetries and autonomic arousal \citep{bradley_1993_emotion, posner_2005_the}. To operationalize this framework for evaluating the emotional intensity of music, two complementary computational strategies are proposed.

\noindent\textbf{Emotional Deviation from Neutral (Valence--Arousal Distance).} Emotional intensity in music can be quantified by measuring the deviation of a piece from emotional neutrality in valence--arousal space. By segmenting a track into $T$ short time windows and somehow estimating valence $v_t$ and arousal $a_t$ per segment, it is possible to calculate the Euclidean distance from the neutral point $(0,0)$. The \textit{\textcolor{metric}{mean emotional distance}} across the track, $\mu_{ES}$, reflects the overall emotional charge, while its standard deviation, the \textit{\textcolor{metric}{emotional variability}} $\sigma_{ES}$, reveals expressivity.

\begin{equation}
\mu_{ES} = \frac{1}{T} \sum_{t=1}^{T} \sqrt{v_t^2 + a_t^2}
\end{equation}
\begin{equation}
\sigma_{ES} = \sqrt{ \frac{1}{T} \sum_{t=1}^{T} \left( \sqrt{v_t^2 + a_t^2} - \mu_{ES} \right)^2 }
\end{equation}

\noindent\textbf{Complexity of the Emotional Trajectory over Time.} Beyond static intensity, emotional content can be conceptualized as the dynamic evolution of valence and arousal over time. Music perceived as highly emotional often exhibits fluctuations, contrasts, and emotional arcs, which musicians refer to as ‘nuance’. To capture this, the rate of change in valence and arousal values across time are computed, denoted $\Delta v_t$ and $\Delta a_t$. The sum of their absolute differences reflects \textit{\textcolor{metric}{emotional volatility}}. Additionally, the \textit{\textcolor{metric}{emotional inflection count}}, i.e., counting inflection points (where the emotional curve changes direction), can capture complexity.

\begin{equation}
\text{Emotional Volatility} = \sum_{t=2}^{T} \left( \left| \Delta v_t \right| + \left| \Delta a_t \right| \right)
\end{equation}
\begin{equation}
\Delta v_t = v_t - v_{t-1}, \qquad \Delta a_t = a_t - a_{t-1}
\end{equation}
\begin{equation}
\text{Emotional Inflection Count} = \text{Number of sign changes in } \Delta v_t \text{ or } \Delta a_t
\end{equation}
The introduced metrics rely heavily on the accuracy of valence and arousal estimations, which are empirically hard to derive, as demonstrated by \citep{yang_2008_a}, who used support vector regression and achieved limited explained variance ($R^2 = 0.58$ for arousal, $0.28$ for valence). Furthermore, the reliance on subjective human annotators to generate the training data used in such studies challenges the goal of evaluating value objectively \citep{grekow_2016_music, yang_2008_a}.

\subsection{Social Validation Through Lyrics as Conceptual Basis for M-value}

When present, lyrics play a key role in shaping the listener’s valuation of a song. They convey emotions, narratives, and cultural messages, often reflecting universal themes such as love, which allow the listener to feel understood. This phenomenon of social validation is further catalyzed when lyrics align with the listener’s personal background. To assess the value of lyrics, Natural Language Processing (NLP) techniques could offer promising tools, as demonstrated in \citep{garg_2025_on}, which represents lyrics as graphs and uses LLMs to trace the evolution of themes and sentiments across time, revealing how lyrics shift with sociocultural context. In reality, the true impact of lyrics depends on how much a listener pays attention to them during listening and the listener's background. For example, someone listening to jazz in a restaurant may disregard lyrical content, while a listener engaging with indie rock in a reflective setting may place high value on the message, which they might value or devalue based on their background.

\subsection{Limitations of Universalization in M-value}

Taking a step back, M-value is inexorably the most subjective component of M-Creativity, and arguably, the most important. In trying to measure M-Creativity, this paper assumed that a single definition of M-value holds true across demographic groups of listeners (PME Assumption). However, some features of music can, at the same time, be highly valued by the listener of a group $A$, while being despised by the listener of a group $B$. An example of this clash is the appreciation of a distortion effect in metal music, which is, broadly speaking, devalued in classical music. In practice, reinforcement learning techniques \citep{sutton_1998_reinforcement} are often the state-of-the-art solutions for quantifying a listener’s valuation of music (M-value). Companies such as Spotify \citep{spotify_2025_spotify} and Deezer \citep{deezer_2025} have invested significant research efforts in building recommendation systems that, over time, leverage individual listener data to model personalized music valuation, avoiding reliance on broad demographic groupings. An example of metrics used to perform such valuations is the ACT-R framework applied in \citep{reiter-haas_2021_predicting}, where components such as base-level (recency and frequency of exposure), spreading (co-occurrence of tracks), and valuation (familiarity and preference reinforcement) are combined to predict relistening behavior, effectively modeling how individual listeners assign value to music through cognitive processes. Google recently introduced MusicRL \citep{cideron_2024_musicrl}, a reinforcement learning-based system for text-to-music generation, which fine-tunes its MusicLM \citep{agostinelli_2023_musiclm} model using both automatic reward functions and large-scale human feedback \citep{cideron_2024_musicrl}. By leveraging Reinforcement Learning from Human Feedback (RLHF), MusicRL optimizes for text adherence, audio quality, and overall listener preference. Crucially, the feedback system can be tailored to specific users, enabling the model to align more closely with individual user valuations rather than generalized listener segments. According to the definition of M-Creativity provided here, such a system would theoretically yield higher listener-perceived value and therefore exhibit greater overall M-Creativity.

\section{M-surprise} \label{sec:msurprise}

According to \citet{boden_2004_the}, the term ``surprise'' has three distinct meanings. Firstly, something can be surprising because it is unfamiliar or unlikely \citep{boden_2004_the}. This kind of surprise goes against probabilities and the founding principle of most music generation algorithms, since the majority of those aim to minimize the artifact’s reconstruction loss \citep{ji_2020_a}. Maximizing M-surprise involves maximizing the unexpectedness of (a) specific musical events or (b) the overall music piece. This kind of surprise occurs when, for instance, a strange chord is being played in a frequently used progression. In a standard V–I (dominant to tonic) cadence, the progression \textit{C–G–Am–F–G–C} establishes a strong tonal center in C major. However, if we replace the expected C major resolution with A minor instead: \textit{C–G–Am–F–G–Am}, this creates a deceptive cadence, where the listener expects a resolution to the tonic (C) but instead gets an unexpected minor chord (Am), which generates surprise. Let us name this type of surprise the \textit{probabilistic surprise (P-s)}, as the essence of such a phenomenon is itself defined using the theoretical idea of minimizing the expectation. It is important to clearly distinguish P-surprise from novelty, as the two concepts are often conflated. In general, if a piece is novel, it is likelier to induce some level of P-surprise. While novelty can contribute to P-surprise, they are not the same: a highly novel piece may still have low internal surprise if its structure is consistent with itself over time.

Secondly, another kind of surprise effect arises when an unexpected idea aligns with an existing way of thinking that you (the surprised) already had \citep{boden_2004_the}. Hence, the surprise comes not from the idea itself being entirely new but from the realization that it naturally fits within a familiar framework, which is something previously unrecognized. This kind of surprise is much harder to evaluate quantitatively as it is a cognitive phenomenon. Let us name this type of surprise the \textit{hidden-content surprise (HC-s)}.

Lastly, surprise occurs when an idea seems utterly impossible, yet it emerges and even inspires other once-unthinkable ideas by using another set of generative principles \citep{boden_2004_the}. It forces a reassessment of what is conceivable and a revision of the current rules. Since this type of surprise is not straightforward to conceive, let us introduce a toy example in Mathematics. The discovery of non-Euclidean geometry overturned the assumption that, through a point not on a given line, exactly one parallel to that line can be drawn. This kind of surprise can be interpreted as a hyperpolation of the known conceptual space stored in the listener’s brain \citep{ord2024hyperpolation}. In fact, such a type of surprise is tied to Boden’s concept of transformational creativity, which is the highest and rarest degree of creativity one observes. This is referred to as \textit{paradigm-shifting surprise (PS-s)}.

\subsection{Probabilistic Surprise (P-surprise)}

When it comes to P-surprise (P-s), using a structural definition of music is convenient as it allows us to distinguish slightly different yet similar phenomena. These are Melodic, Harmonic, Timbral, and Rhythmic P-surprises. Indeed, it is conceivable for a song to have highly P-surprising harmonies and yet a very basic rhythm, which would be familiar to the listener. Altogether, these components make a reasonable heuristic to measure the overall level of P-surprise in a piece of music.

\subsubsection{Measuring Melodic and Harmonic P-surprise in AI-Music Artifacts}

To capture melodic and harmonic P-surprise beyond individual musical events, two complementary macroscopic metrics are introduced: \textit{Mean IC (Information Density)} and \textit{Entropy of Musical Events}. While Mean IC captures the average contextual surprise of realized events, entropy captures their variability, with artifact-derived entropy measuring the diversity of events within a composition and model-derived entropy measuring the generative model's predictive uncertainty.

\paragraph{Melodic/Harmonic Information Content}

A well-established method for measuring melodic and harmonic P-surprise is based on Shannon’s Information Theory \citep{shannon_1948_a}. In this approach, surprise relates directly to the predictability of a musical event (note, chord) given its preceding context, leading to \textit{\textcolor{metric}{melodic}} and \textit{\textcolor{metric}{harmonic information content}}. An event with low probability carries high Information Content (IC) \citep{cohen_1962_information}, signifying greater surprise, whereas highly predictable events yield low IC:
\begin{equation}\label{eq:ic}
IC\left(x_t\right) = -\log{P}\left(x_t \mid x_{<t}\right)
\end{equation}
Probabilities are estimated from listeners’ Prior Musical Experience (PME), typically modeled using frequency-based N-gram methods \citep{jurafsky_2026_speech}:
\begin{equation}\label{eq:ngram}
P\left(x_t \mid x_{t-n}, \ldots, x_{t-1}\right) = \frac{\mathrm{Count} \left(x_{t-n}, \ldots, x_t\right)}{\mathrm{Count}\left(x_{t-n}, \ldots, x_{t-1}\right)}
\end{equation}
N-gram models are computationally efficient and interpretable but suffer from limited context length and data sparsity \citep{jurafsky_2026_speech}. More sophisticated sequence models (e.g., Transformers, LSTMs) capture longer-range dependencies and subtle musical patterns but often at the expense of interpretability.
Empirical evidence supports IC’s effectiveness in measuring P-surprise in AI-generated music. \citet{bjare2025musicalsurprisal} found that the IC of a Generative Infinite-Vocabulary Transformer trained over latent audio representations correlates with EEG responses of listeners to songs.

\paragraph{Melodic/Harmonic Mean IC (Information Density)} 

The \textit{\textcolor{metric}{mean melodic}} or \textit{\textcolor{metric}{harmonic IC}} measures the average information content across musical segments of notes/chords, evaluating overall surprise density:
\begin{equation}
IC_{\mathrm{mean}} = \frac{1}{N} \sum_{t=1}^{N} IC\left(x_t\right)
\end{equation}
\paragraph{Entropy of Musical Events} 

The \textit{\textcolor{metric}{melodic}} or \textit{\textcolor{metric}{harmonic entropy}} quantifies the inherent unpredictability of a distribution over musical events. These two variants specialize the Shannon entropy
\begin{equation}\label{eq:entropy}
H\left(X\right) = -\sum_{x}{P\left(x\right)\log{P}\left(x\right)},
\end{equation}

\noindent and differ only in \textit{which} distribution $P$ plays the role of $P(x)$, reusing the probabilities already introduced for IC in Eqs.~\eqref{eq:ic}--\eqref{eq:ngram}.

\noindent \textbf{Artifact-derived Entropy of Musical Events.} In this case, $P(x)=\hat{P}(x)$ is the empirical (maximum-likelihood) frequency of event $x$ \textit{within a single composition}---the unigram marginal of the event counts of Eq.~\eqref{eq:ngram}. It reflects the internal variability of that one artifact:
    \begin{equation}
    H_{\mathrm{art}}(X) = -\sum_{x} \hat{P}(x)\,\log \hat{P}(x).
    \end{equation}

\noindent \textbf{Model-derived Entropy of Musical Events.} In this case, $P$ is the generative model’s own predictive distribution $P(x_t \mid x_{<t})$ from Eq.~\eqref{eq:ic}, averaged over the sequence. This is the model’s expected information content---the mean of the per-event IC of Eq.~\eqref{eq:ic}---and reflects its potential for generating diverse and unexpected melodies and harmonies:
    \begin{equation}
    H_{\mathrm{model}} = \frac{1}{N}\sum_{t=1}^{N}\Big(-\sum_{x} P(x \mid x_{<t})\,\log P(x \mid x_{<t})\Big) = \mathbb{E}\!\left[IC(x_t)\right].
    \end{equation}

\subsubsection{Measuring Timbral P-surprise in Music Artifacts}

While melodic and harmonic P-surprise are measured through symbolic predictability, timbral P-surprise concerns the tonal color or texture of sound. Timbre distinguishes instruments and sonic textures \citep{McAdams2009}, and its surprise arises when an unexpected shift in tone occurs. This can manifest either locally (e.g., a distorted guitar entering a classical piano piece) or globally, with frequent or intense timbral changes throughout a composition (e.g., Queen's \textit{Bohemian Rhapsody} \citep{queen1975bohemian}, which moves from a cappella vocals to piano ballad, operatic chorus, and hard rock within a single track). Given the focus on M-Creativity as a perceptual phenomenon, timbral surprise is best assessed in a representation aligned with human auditory perception. For this reason, the Mel-Spectrogram is used, which scales frequency bins to better match the resolution of human hearing.

\paragraph{Spectral Flux (SF)}

A key metric for quantifying timbral change is \textit{\textcolor{metric}{Spectral Flux}} (SF), which measures the degree of spectral variation between consecutive time frames \citep{tzanetakis_2002_musical,giannakopoulos_2014_introduction}:
\begin{equation}
SF(t) = \sum_{k} \left( \left|S_{\mathrm{mel}}(k,t)\right| - \left|S_{\mathrm{mel}}(k,t-1)\right| \right)^2
\end{equation}
\noindent where \(S_{\mathrm{mel}}(k, t)\) is the Mel-spectrogram magnitude at frequency bin \(k\) and time \(t\). High SF indicates sharp timbral changes (e.g., new instruments or abrupt attacks), while low SF reflects smooth, continuous transitions. To measure Timbral P-surprise, SF values are interpreted relative to the Prior Musical Experience (PME) of the listener. First, a baseline average SF is computed from a PME corpus reflecting the listener’s typical listening background. Then, deviations in the target musical artifact are identified and normalized. Frequent or large deviations from PME-defined expectations signal timbral surprise.

\paragraph{Z-Score Based P-surprise} 

Two complementary methods for detecting P-surprise based on spectral features are proposed, capturing unexpected changes in the spectral characteristics of musical artifacts at different levels of abstraction.

    \noindent \textbf{Local Surprise Detection.} For each time frame \(t\), it computes the \textit{\textcolor{metric}{Z-score (local)}}:
    \begin{equation}
    Z_{sf}(t) = \frac{SF(t) - \mu_{\mathrm{PME}}}{\sigma_{\mathrm{PME}}}
    \end{equation}
    \noindent where \(\mu_{\mathrm{PME}}\) is the mean SF from the PME corpus and \(\sigma_{\mathrm{PME}}\) is the standard deviation of SF in the PME. If \(Z_{sf}(t) \gg 0\), the frame is unexpectedly timbrally varied. One issue with such an approach is that it compares a mean against a single frame.

    \noindent \textbf{Global (Mean) Surprise Detection.} To get a fairer comparison, it is possible to compare the mean SF from the PME and the mean SF of the music artifact or all artifacts generated by the AI model, i.e., the  \textit{\textcolor{metric}{Z-score (global)}}:
    \begin{equation}
    \bar{Z}_{sf} = \frac{\mu_{\mathrm{artifact}} - \mu_{\mathrm{PME}}}{\sigma_{\mathrm{PME}}}
    \end{equation}
    \noindent where \(\mu_{\mathrm{artifact}}\) is the mean SF of the target piece or of all outputs from the AI model, while \(\mu_{\mathrm{PME}}\) and \(\sigma_{\mathrm{PME}}\) are the mean and standard deviation of SF from the PME.
Partitioning the PME by genre can further refine the above techniques by, for instance, calculating \(\bar{Z}_{sf}\) for each group of music genre and therefore having contextualized measurements, conscious that different genres pertain to different sets of timbral norms. 

\paragraph{Entropy of Spectral Flux} 
In analogy to melodic/harmonic surprise metrics, it is possible to compute the entropy of SF values to assess the overall unpredictability or variability of timbre, i.e., the \textit{\textcolor{metric}{timbral entropy}}. There are two applications that are summarized below.

\noindent \textbf{Artifact-derived Timbral Entropy.} This is computed from the distribution of SF values within a single musical piece. High entropy reflects erratic or diverse timbral changes, suggesting a chaotic and surprising texture.

\noindent \textbf{Model-derived Timbral Entropy.} This is computed from SF values across all outputs generated by a music model. A high value suggests the model can produce varied timbral textures; low entropy suggests uniformity.

\subsubsection{Measuring Rhythmic P-surprise in a Musical Artifact}

Rhythm, the temporal organization of musical events, plays a central role in shaping listener expectations and perceptual engagement \citep{Thaut2014}. In many cases, listeners identify a song more easily by its rhythm than by its harmony. From a probabilistic perspective, rhythmic P-surprise emerges when the timing of musical events deviates from patterns internalized via Prior Musical Experience (PME). These deviations may include syncopation, irregular groupings (e.g., quintuplets in regular meter), unexpected silences, or expressive timing shifts. Rhythmic expectation can be modeled using inter-onset intervals (IOIs), which represent the time between successive rhythmic events:
\begin{equation}
\mathrm{IOI}_k = t_k - t_{k-1}
\end{equation}
A rhythm stream is thus defined as a sequence of IOIs. Rhythmic P-surprise can be quantified by computing how probable each IOI is under the listener’s PME. However, rhythm is inherently polyphonic; multiple rhythmic layers (e.g., drums, bass, vocals) coexist and interact perceptually. In genres like rap, the drum pattern (the grid) and the vocal line (flow) may diverge significantly, generating perceived groove and tension. Decomposing these overlapping patterns is straightforward with symbolic data (e.g., MIDI), where instruments are stored in separate channels. In such cases, rhythmic surprise metrics (e.g., IC, entropy) can be computed per component. This task becomes more complex with real-world audio formats (e.g., MP3, WAV), where individual tracks are merged. In this case, rhythm cannot be directly analyzed at the component level without preprocessing. To address this, audio-based source separation algorithms such as Demucs \citep{defossez_2021_hybrid} and Spleeter \citep{hennequin_2020_spleeter} can be used to isolate instrument streams from mixed audio. Once separated, IOI patterns can be extracted and analyzed per stream. Alternatively, rhythm can be modeled without decomposition, using spectral rhythm patterns \citep{peeters_2005_rhythm}. These methods derive rhythmic structure directly from audio by analyzing periodicities in onset energy curves using techniques such as the Discrete Fourier Transform (DFT) \citep{palani_2022_discrete} and the Auto-Correlation Function (ACF) \citep{box_2015_time}. DFT transforms temporal signals into frequency components, while ACF identifies repeating structures over time. These methods are especially useful when symbolic or separated data is unavailable. However, they trade off interpretability and precision, particularly for voice-specific rhythmic behavior. Because of the framework's emphasis on explainability \citep{dwivedi_2023_explainable}, this paper focuses on decomposition-based methods.

\paragraph{Information Theoretic Lens}

\noindent \textbf{Information Content of IOIs.}  The  \textit{\textcolor{metric}{IOIs information content}} captures the rhythmic P-surprise at the micro level by estimating how unexpected a specific IOI is within a given rhythmic context:
\begin{equation}
IC(\mathrm{IOI}_k) = -\log P_\varepsilon(\mathrm{IOI}_k \mid \text{context})
\end{equation}
where \(P_\varepsilon\) incorporates a microtiming tolerance \citep{butterfield_2010_participatory}:
\begin{equation}
P_\varepsilon(\mathrm{IOI}_k \mid \text{context}) = \int_{\mathrm{IOI}_k - \varepsilon}^{\mathrm{IOI}_k + \varepsilon} p(\tau \mid \text{context}) \, d\tau
\end{equation}
This reflects moments where a performer plays ahead or behind the grid in ways that deviate from prior expectations, creating perceptible rhythmic P-surprise. As with melodic and harmonic P-surprise, the segmental rhythmic P-surprise can be computed using the mean IC (Information Density).

\noindent \textbf{Entropy of IOIs.}  The  \textit{\textcolor{metric}{rhythmic entropy}} captures the potential for surprise by measuring the variability of IOI durations:

\begin{equation}
H(\mathrm{IOI}) = -\sum P(\mathrm{IOI}_i) \log P(\mathrm{IOI}_i)
\end{equation}

It is possible to consider \textit{artifact-derived entropy}, computed from the empirical distribution of IOI values within a music piece. A rhythmically simple piece with repetitive IOIs (e.g., a metronomic techno beat) yields low entropy. Alternatively, it is possible to consider \textit{model-derived entropy}, which reflects a model’s capacity to produce rhythmically diverse outputs. A high value suggests the model can generate a wide range of IOIs, potentially increasing surprise.

Entropy of IOIs can be used alongside IC to distinguish between realized surprise (IC) and latent potential for surprise (entropy). A rhythmically unpredictable piece may not always be surprising, but it holds the capacity to be, depending on the listener's PME.

\paragraph{Musicological Lens}

\noindent \textbf{Tempo Deviation Surprise (TDS).} Using beat tracking algorithms, the tempo trajectory across a piece can be extracted. Sudden deviations or expressive shifts (e.g., rubato, swing) are detected by computing the derivative of the tempo curve and comparing it with expectations from the PME. Let \(T(t)\) denote the instantaneous tempo (in BPM) at time \(t\), and its derivative:
\begin{equation}
\Delta T(t) = \frac{dT(t)}{dt}
\end{equation}
Then, the \textit{\textcolor{metric}{tempo deviation surprise}} is calculated as:
\begin{equation}
S_{\mathrm{tempo}}(t) = \left| \frac{ \Delta T(t) - \mu_{\Delta T(t)} }{ \sigma_{\Delta T(t)} } \right|
\end{equation}
High \(S_{\mathrm{tempo}}\) values indicate surprising tempo modulations. This reflects another kind of rhythmic P-surprise rooted in expressive nuance, and the tempo trajectory itself can be extracted with modern beat- and downbeat-tracking methods such as that of \citet{deng_2024_efficient}, which combines Temporal Convolutional Networks with self-supervised DistilHuBERT features.

\noindent \textbf{Cross-Stream Rhythmic Dissonance.} 
Cross-Stream Rhythmic Dissonance compares IOI distributions across streams (e.g., vocals vs. percussion). Sharp divergence between patterns suggests intentional disruption of rhythmic alignment. The \textit{\textcolor{metric}{IOI distribution divergence}} can be operationalized via Jensen–Shannon Divergence (JSD) between the normalized IOI histograms:
\begin{equation} 
\mathrm{JSD}(P \parallel Q) = \frac{1}{2} D_{\mathrm{KL}}(P \parallel M) + \frac{1}{2} D_{\mathrm{KL}}(Q \parallel M)
\end{equation}
where \(M = \frac{1}{2}(P + Q)\), and \(D_{\mathrm{KL}}\) is the Kullback–Leibler divergence. A high JSD between, for example, the vocal and drum tracks indicates rhythmically dissonant layering, often perceived as surprising or creative interplay, though it can also lead to output of little value and diminish the M-value score.

\subsection{Hidden-Content Surprise (HC-surprise)}

HC-surprise is a much harder concept to evaluate quantitatively, as it is a purely cognitive phenomenon. It would require us to be able to understand how the brain organizes knowledge (Prior Musical Experiences) and processes new stimuli (unexpected music artifacts). Several paradigms suggest that the brain behaves as a predictive machine, where it continuously updates its expectations to minimize surprise \citep{clark_2013_whatever, levitin_2006_this, patel_2008_music}, and engages in Gestalt reconfiguration, where existing knowledge is suddenly reorganized into a coherent structure, leading to an ‘aha!’ moment of realization in problem-solving \citep{wertheimer_1938_gestalt}. Assuming the brain indeed behaves as such (and that the ‘aha’ moment corresponds to the second kind of surprise described by \citet{boden_2004_the}) would suggest that HC-surprise arises from restructuring musical knowledge with respect to the newly processed music artifact into a coherent whole. The HC-surprise phenomenon can manifest across various musical dimensions. For example, a listener may suddenly recognize that an unexpected word in the lyrics reflects their own thoughts, deepening their connection to the music. Instrumentally, HC-surprise might arise when a surprising rhythmic shift resonates with patterns the listener already knows from another genre (e.g., moving from pop-rock to reggae in a song). This effect can be especially strong for musicians, who may have previously imagined or explored similar transitions in their own creative work.

Measuring HC-surprise empirically is difficult, as it would need 1) A model of how the listener structures musical knowledge; 2) A way to detect when their internal model is reorganized; and 3) A definition of what counts as ``coherent reorganization''.
Another way one can conceive HC-surprise is through the Bayesian lens. If we think of the listener’s brain as a sort of model, or rather, think of the model as an approximation of the listener’s brain (PME assumption), HC-s could be reframed as a kind of a posteriori surprise, defined not only by an initial violation of expectations (high P-surprise), but also by a minimization of the posterior divergence, which corresponds to how much your beliefs (probability distribution) change after seeing new evidence. HC-s would occur in a two-step process. Firstly, a listener finds some musical elements or an entire musical piece highly P-s. Secondly, thinking of it in retrospect, the listener no longer finds it so surprising, as it turns out to fit well with their internal model. As can be sensed here, this means that the listener would be acting quite paradoxically. This ability to adopt various strategies of thinking, which can in some situations lead to paradox, is deeply human and very hard to instill in machines.
Nonetheless, ensembling different models sequentially, with different ‘ways of thinking’ (architectures) and ‘knowledge’ (data) at the meta level, could be seen as mimicking humans’ ability to exhibit different reasonings and have second thoughts, which can be paradoxical. A highly P-s musical stimulus $x$ is first evaluated by Model 1, which registers a large belief update. Then, Model 2 reinterprets $x$ with minimal belief change, recognizing it as coherent after all.

An information-content reading offers a complementary, and arguably cleaner, criterion for the ‘aha’ moment. Let $\mathrm{IC}(K)$ denote the information content required to encode a listener’s musical knowledge $K$, and let $K' = K \cup \{x\}$ be that knowledge after the surprising artifact $x$ has been integrated and its consequences propagated. A genuine ‘aha’ is then signaled not merely by a large initial belief update, but by a subsequent \textit{reduction} in description length,
\begin{equation}
\mathrm{IC}(K') < \mathrm{IC}(K),
\end{equation}
that is, the surprising piece ultimately lets the whole body of knowledge be re-encoded more compactly, having revealed latent structure rather than merely added noise. This equates insight with a gain in compression---a view with precedent in the information-theoretic account of interestingness and creativity as compression progress \citep{schmidhuber_2010_formal}---and dovetails with the Bayesian formulation below: the KL update of Eq.~\eqref{eq:bayes-surprise} captures the initial ``surprise'' step, while the drop in $\mathrm{IC}$ captures the subsequent ``re-arrangement into a more structured whole''. This is worth stating explicitly, even though it remains as hard to operationalize as the Bayesian version, since it presupposes an explicit encoder of a listener’s musical knowledge.

Formally, let each model $M$ maintain beliefs over latent musical structures $\theta$. The surprise experienced by model $M$ upon hearing a musical input $x$ is the Bayesian surprise \citep{itti_2009_bayesian}, defined as the Kullback-Leibler (KL) divergence between the posterior and prior distributions:
\begin{equation}\label{eq:bayes-surprise}
S_M^{\mathrm{Bayes}}(x) = D_{\mathrm{KL}}(P^M(\theta\mid x)\parallel P^M(\theta))
\end{equation}
This measures how much the model’s belief over $\theta$ shifts after observing $x$.

Then, the \textit{\textcolor{metric}{Bayesian HC-surprise}} is defined as the contrast in belief updates across two models with different reasoning strategies:
\begin{equation}
\mathrm{HC\text{-}s}(x) = S_1^{\mathrm{Bayes}}(x) - S_2^{\mathrm{Bayes}}(x)
\end{equation}
If $\mathrm{HC\text{-}s}(x) \gg 0$: the input is initially surprising but later becomes retrospectively coherent. In more complex systems with \( n \) models \(\{M_1, M_2, \dots, M_n\}\), each weighted by its cognitive relevance \( w_i \) (where \( \sum w_i = 1 \)), a general form of Bayesian HC-surprise can be seen as:
\begin{equation}
\mathrm{HC\text{-}s}(x) = \sum_{i} w_i \cdot D_{\mathrm{KL}}\left(P^{(i)}(\theta \mid x) \,\|\, P^{(i)}(\theta)\right)
\end{equation}

\begin{figure}[t]
    \centering
    \includegraphics[width=\textwidth]{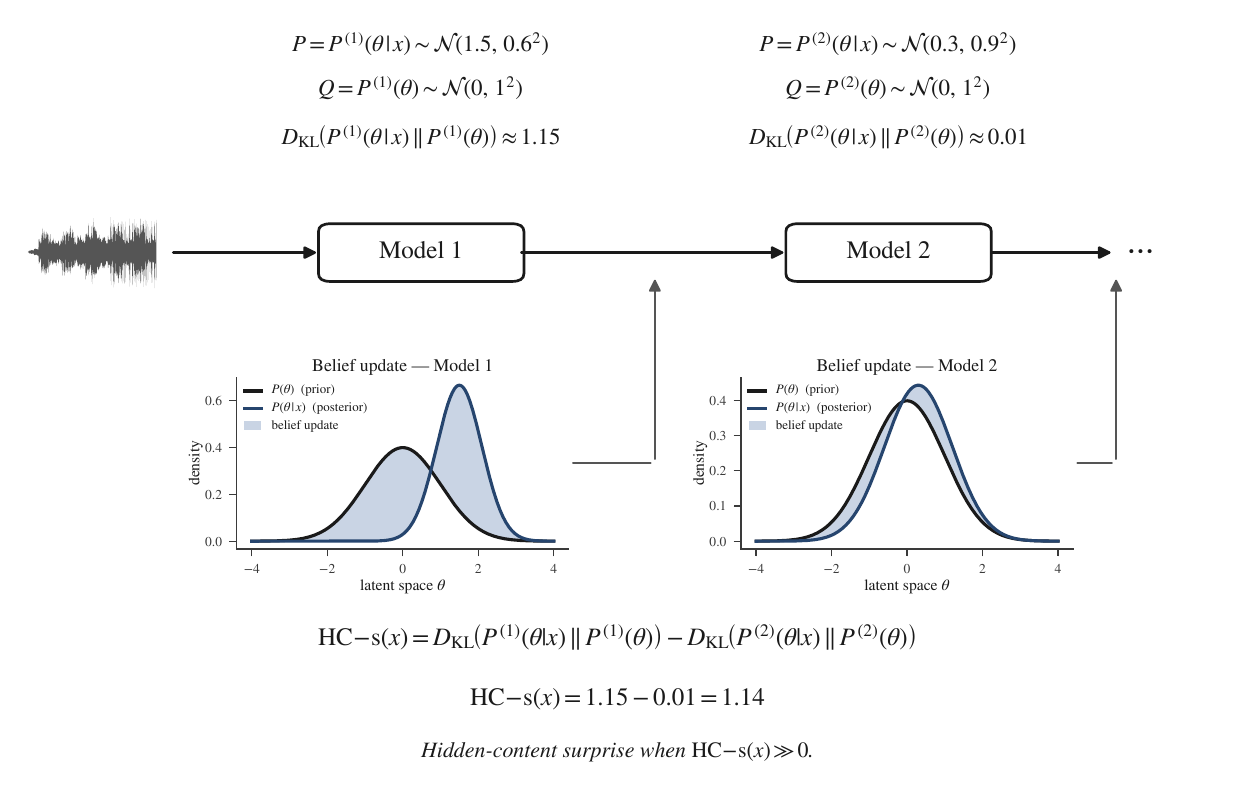}
    \caption{Illustration of HC-surprise computed from Bayesian model divergences.}
    \label{fig:hcs}
\end{figure}

This formulation, illustrated in Figure~\ref{fig:hcs}, models human-like layered reasoning and provides ideas for future architectures that could unlock the potential for paradoxical conclusions in AI music generation systems. Empirically assessing HC-surprise would require AI models capable of dynamically updating their internal distributions and reasoning in multiple, potentially contradictory, ways. Yet, current generative architectures lack the capacity to simulate such higher-order cognitive restructuring. As a result, evaluating HC-surprise in artifacts produced by modern generative AI music systems remains both conceptually and technically unfeasible.

\subsection{Paradigm-Shifting Surprise (PS-surprise)}
Paradigm-Shifting Surprise (PS-surprise) refers to the most radical form of musical surprise, i.e., when an artifact expands the conceptual space of music itself. It corresponds to Boden's idea of transformational creativity \citep{boden_2004_the}, where the generative rules of a system are redefined rather than merely subverted. The shift from tonal harmony to atonality in 20th-century classical music can be seen as an example of PS-s in music. Such transformations can be described as cases of hyperpolation \citep{ord2024hyperpolation}, the expansion of conceptual space through the creation of entirely new musical dimensions. For example, the invention of an instrument that supports a new tuning system and distinct timbre could generate works that existing instruments cannot replicate. If listeners accept these artifacts as music, then the conceptual space of music has been irrevocably redrawn.

\subsubsection{Impossibility of Direct Measurement}
PS-surprise fundamentally challenges the assumptions of canonical or pre-trained models, which are limited to existing musical structures. A truly PS-surprising piece introduces dimensions not present in training data, making it unrepresentable within the model’s conceptual space. For example, a model trained only on 2D inputs—say, voice and piano—cannot generate or meaningfully process 3D musical artifacts that include an added guitar stream.

Another difficulty in operationalizing PS-surprise lies in the necessity of knowing the rules that define a given musical system. While in certain well-defined genres, such as classical tonal music, these rules are relatively explicit, many contemporary genres like pop-rock operate under fluid and evolving conventions. In such cases, the boundaries of what counts as a legitimate musical structure are less clearly defined, making it harder to assess when a shift has occurred.

Furthermore, even if we could delineate the rules of a given style and recognize an artifact that clearly violates them, the recognition of PS-surprise is not instantaneous. From the perspective of cultural evolution theory \citep{henrich_2015_the}, PS-surprising music can be seen as ‘mutations’ in the cultural gene pool, ideas that initially lack ‘cultural fitness’ but, under shifting conditions, may reshape musical norms. If adopted, they open up a new branch in the evolutionary tree of music.

\subsubsection{Properties of PS-s Music}
There are three necessary (but not sufficient) conditions:
\begin{itemize}
    \item \textit{Exceptionally High M-novelty}: The artifact diverges radically from all previously known human or AI-generated music. It occupies a previously uncharted region of the musical conceptual space.
    \item \textit{Exceptionally High P-surprise}: Listeners experience extreme unexpectedness emanating from new musical dimensions.
    \item \textit{Irrepresentability}: Canonical models fail to classify or interpret the artifact meaningfully.
\end{itemize}

In essence, it is perhaps because of its meta-existence in music that we struggle to find a proper way to measure PS-s in a music artifact. Indeed, discerning whether some music artifact is musically PS-s would require knowing the ultimate rules of music, which are not even set clearly among humans, and having a meta-view on music, which is hard to conceptualize. 

Taking a step back, musical surprise may extend beyond P-s, HC-s, and PS-s. While Boden’s framework helps distinguish forms of creativity, it remains difficult to fully map onto music. In particular, PS-surprise, rooted in the abstract notion of transformational creativity, often aligns better with mathematical or conceptual systems.

\section{M-novelty} \label{sec:mnovelty}

M-novelty is the degree to which music generated by an AI system deviates from its reference data. When measured against the model’s training set, this reflects psychological novelty—something new to the system itself. This aligns with P-creativity, where novelty is defined from the creator’s perspective rather than the observer’s \citep{boden_2009_computer, barto_2013_novelty}. A composition is M-novel if it lies outside the learned distribution and is thus unfamiliar to the model internally. This provides a quantifiable proxy for creativity: since the system only learns from its training data, deviations represent interpolation or extrapolation beyond prior exposure—forms of creative behavior \citep{ord2024hyperpolation}. The same logic extends to historical novelty (M-H-novelty): by changing the reference point from the training set to a broader musical corpus, we assess whether the generated piece is novel in relation to known musical works rather than just the model’s experience.

\subsection{Embedding-Based M-novelty}

To operationalize M-novelty, an embedding-based distance can be used.
Let $\mathbf{v}_{\mathrm{gen}}$ be the embedding of the generated piece,
and $\{\mathbf{v}_1,\ldots,\mathbf{v}_N\}$ be the embeddings of training pieces.
Cosine distance is chosen over Euclidean distance, as embedding vector magnitudes often lack clear musical interpretation (e.g., in high-dimensional spaces such as MERT \citep{li_2024_mert}):
\begin{equation}
\cos\!\big(\mathbf{v}_{\mathrm{gen}}, \mathbf{v}_i\big)
  = \frac{\mathbf{v}_{\mathrm{gen}}\cdot \mathbf{v}_i}
         {\lVert \mathbf{v}_{\mathrm{gen}}\rVert \,\lVert \mathbf{v}_i\rVert}
\end{equation}

The \textit{\textcolor{metric}{average}} or \textit{\textcolor{metric}{minimum cosine distance}} between $\mathbf{v}_{\mathrm{gen}}$ and the training set provides a scalar estimate of M-novelty:
\begin{equation}
d_{\mathrm{avg}}\!\big(\mathbf{v}_{\mathrm{gen}}\big)
  = \frac{1}{N}\sum_{i=1}^N \Big(1 - \cos\!\big(\mathbf{v}_{\mathrm{gen}}, \mathbf{v}_i\big)\Big)
\end{equation}
\begin{equation}
d_{\min}\!\big(\mathbf{v}_{\mathrm{gen}}\big)
  = \min_{1 \le i \le N} \Big(1 - \cos\!\big(\mathbf{v}_{\mathrm{gen}}, \mathbf{v}_i\big)\Big)
\end{equation}

Although statistical divergence metrics like KL divergence are often used in related contexts, they are deliberately avoided here. As argued by \citet{barto_2013_novelty}, such measures quantify unexpectedness under a reference distribution, which is better linked to surprise than to novelty. To prevent conceptual overlap between these dimensions, this paper instead adopts direct embedding-space comparisons that better reflect the generative system’s own knowledge boundary.

Two caveats attach to any embedding-based novelty measure. First, the measure is only as meaningful as the embedder: there is no canonical, ``optimal'' music encoder, and novelty scores are therefore relative to the representation chosen (e.g., MERT \citep{li_2024_mert} versus a joint audio--text space such as CLAP). Comparisons across systems or studies are meaningful only when the embedder is held fixed. Second, cosine distance carries its own assumptions. By discarding vector magnitude, it treats direction as the sole carrier of meaning, which need not hold when magnitude encodes salience or intensity; moreover, contextual embedding spaces are known to be \textit{anisotropic}, occupying a narrow cone that systematically inflates pairwise cosine similarity and compresses the usable dynamic range of the measure \citep{ethayarajh_2019_how}. Cosine distance is thus a convenient default rather than a universally appropriate metric, and reported novelty values should be read as embedder- and metric-relative.

These limitations motivate complementary, non-embedding notions of novelty. One alternative decomposes an artifact into a set of discrete, interpretable properties---for instance, the presence of a key modulation, a particular meter, or a given instrument class---and compares artifacts on these attributes in a binary fashion, yielding a transparent, if coarser, novelty signal that does not depend on a learned space. Relatedly, because each artifact is decomposed into several musical dimensions (melody, harmony, rhythm, timbre)---as in M-surprise and in the paragraph that follows---a per-dimension novelty vector must be reconciled into a single score. The two natural choices are an \textit{average}, which rewards music that is broadly novel across dimensions, and a \textit{maximum}, which rewards a striking departure along even a single dimension. As these encode different aesthetic priorities, reporting both rather than committing to one is recommended.

Beyond macroscopic M-novelty evaluation, it is possible, akin to what has been proposed in M-surprise, to decompose each artifact into interpretable musical dimensions such as melody, harmony, rhythm, and timbre, and compute the \textit{\textcolor{metric}{decomposed embedding novelty}} per component. This decomposition enhances explainability and allows us to identify which aspects of a composition contribute most to its novelty. For instance, generated music might contain highly novel melodic streams layered up with common rhythmic streams such as a steady 4/4 kick-snare pattern.

\section{Summary of Proposed Metrics}

This paper proposes a set of metrics intended to capture aspects of M-Creativity. Table~\ref{tab:mcreativity-multipage} summarizes these metrics, categorized under M-surprise, M-novelty, and M-value, along with their respective sub-components and measurement methods.

\begingroup
\small
\renewcommand{\arraystretch}{1.3}%
\setlength{\extrarowheight}{2pt}
\begin{longtable}{@{}>{\RaggedRight\arraybackslash}m{\dimexpr 0.25\linewidth-\tabcolsep\relax}
                   >{\centering\arraybackslash}m{\dimexpr 0.47\linewidth-2\tabcolsep\relax}
                   >{\RaggedRight\arraybackslash}m{\dimexpr 0.28\linewidth-\tabcolsep\relax}@{}}
  \caption{Summary of M-Creativity Metrics.}
  \label{tab:mcreativity-multipage}
  \\
  \toprule
  \textbf{Metric} & \textbf{Formula} & \textbf{Interpretation} \\
  \midrule
  \endfirsthead

  \multicolumn{3}{c}{{\bfseries \tablename\ \thetable{} -- continued from previous page}} \\
  \toprule
  \textbf{Metric} & \textbf{Formula} & \textbf{Interpretation} \\
  \midrule
  \endhead

  \midrule \multicolumn{3}{r}{{\textit{Continued on next page}}} \\
  \endfoot

  \bottomrule
  \endlastfoot

  \multicolumn{3}{@{}l}{\textbf{M-value}} \\

  Mean Emotional Distance &
  $\dfrac{1}{T}\sum_{t=1}^{T}\sqrt{v_t^2+a_t^2}$ &
  Average emotional intensity over time \\[5pt]

  Emotional Variability &
  $\sqrt{\dfrac{1}{T}\sum_{t=1}^{T}\left(\sqrt{v_t^2+a_t^2}-\mu_{ES}\right)^2}$ &
  Expressivity: variation in emotional strength \\[5pt]

  Emotional Volatility &
  $\sum_{t=2}^{T}\big(\lvert\Delta v_t\rvert+\lvert\Delta a_t\rvert\big)$ &
  Emotional fluctuation over time \\[5pt]

  Emotional Inflection Count &
  Count of sign changes in $\Delta v_t$ or $\Delta a_t$ &
  Complexity of emotional arc \\[5pt]

  \addlinespace[4pt]
  \multicolumn{3}{@{}l}{\textbf{M-surprise}} \\

  Melodic Information Content (IC) &
  $-\log P(x_t \mid x_{<t})$ &
  Predictive surprise of a melodic note \\[5pt]

  Mean Melodic IC &
  $\dfrac{1}{N}\sum IC(x_t)$ &
  Average surprise across melodic sequence \\[5pt]

  Melodic Entropy &
  $-\sum P(x)\log P(x)$ &
  Variability in melodic distribution \\[5pt]

  Harmonic Information Content (IC) &
  $-\log P(h_t \mid h_{<t})$ &
  Predictive surprise of a chord \\[5pt]

  Mean Harmonic IC &
  $\dfrac{1}{N}\sum IC(h_t)$ &
  Average harmonic surprise \\[5pt]

  Harmonic Entropy &
  $-\sum P(h)\log P(h)$ &
  Variability in chord distribution \\[5pt]

  Spectral Flux (SF) &
  $\sum\nolimits_k \big(\lvert S_{\text{mel}}(k,t)\rvert - \lvert S_{\text{mel}}(k,t-1)\rvert\big)^{2}$ &
  Local timbral change \\[5pt]

  Z-score (Local) &
  $\dfrac{SF(t)-\mu_{\text{PME}}}{\sigma_{\text{PME}}}$ &
  Deviation from mean PME flux \\[5pt]

  Z-score (Global) &
  $\dfrac{\mu_{\text{artifact}}-\mu_{\text{PME}}}{\sigma_{\text{PME}}}$ &
  Overall timbral deviation \\[5pt]

  Timbral Entropy &
  $-\sum P(SF)\log P(SF)$ &
  Timbral variability \\[5pt]

  IOIs Information Content &
  $-\log P_\varepsilon(\mathrm{IOI}_k \mid \text{context})$ &
  Surprise from rhythm timing \\[5pt]

  Mean IOIs IC &
  $\dfrac{1}{N}\sum IC(\mathrm{IOI}_k)$ &
  Average rhythmic surprise \\[5pt]

  Rhythmic Entropy &
  $-\sum P(\mathrm{IOI}_i)\log P(\mathrm{IOI}_i)$ &
  Rhythmic pattern variability \\[5pt]

  Tempo Deviation Surprise &
  $\left\lvert\dfrac{\Delta T(t)-\mu_{\Delta T}}{\sigma_{\Delta T}}\right\rvert$ &
  Surprise from tempo change \\[5pt]

  IOI Distribution Divergence &
  $\tfrac{1}{2}D_{\mathrm{KL}}(P\parallel M)+\tfrac{1}{2}D_{\mathrm{KL}}(Q\parallel M)$,\newline
  $M=\tfrac{1}{2}(P+Q)$ &
  Cross-stream rhythm dissonance \\[5pt]

  Bayesian HC-surprise (two models) &
  $S_1^{\mathrm{Bayes}}(x)-S_2^{\mathrm{Bayes}}(x)$, \newline
  \mbox{$S_i^{\mathrm{Bayes}}(x)=D_{\mathrm{KL}}\big(P^{(i)}(\theta\mid x)\,\|\,P^{(i)}(\theta)\big)$} &
  Cognitive reconfiguration surprise \\[5pt]

  Bayesian HC-surprise (weighted sum) &
  $\sum_{i=1}^{n} w_i\, D_{\mathrm{KL}}\big(P^{(i)}(\theta \mid x)\,\|\,P^{(i)}(\theta)\big)$, \newline
  $\sum_{i=1}^{n} w_i=1,\; w_i\ge 0$ &
  Aggregated belief-update across $n$ models/strategies \\[5pt]

  \addlinespace[4pt]
  \multicolumn{3}{@{}l}{\textbf{M-novelty}} \\

  Embedding Cosine Similarity (basis of the distances below) &
  $\dfrac{v_{\text{gen}}\cdot v_i}{\lVert v_{\text{gen}}\rVert\,\lVert v_i\rVert}$ &
  Similarity vs.\ training set items \\[5pt]

  Average Cosine Distance &
  $\dfrac{1}{N}\sum_{i=1}^{N}\Big(1-\cos\big(v_{\text{gen}},v_i\big)\Big)$ &
  Aggregate novelty vs.\ training corpus (average pairwise cosine distance) \\[5pt]

  Minimum Cosine Distance &
  $\min_{1\le i\le N}\Big(1-\cos\big(v_{\text{gen}},v_i\big)\Big)$ &
  Nearest-neighbor novelty vs.\ training set \\[5pt]

  Decomposed Embedding Novelty &
  Same formulas applied per dimension\slash stream &
  Melody\slash harmony\slash timbre\slash rhythm novelty \\[5pt]
\end{longtable}
\endgroup

\section{Discussion}

This paper has formalized a set of actionable metrics to assess musical creativity in AI-generated compositions with relative objectivity (see Table~\ref{tab:mcreativity-multipage} for a summary). This suggests that it may be possible, at least in principle, to quantify objectively key aspects of musical creativity and that Margaret Boden’s framework \citep{boden_2004_the} provides a strong foundation for doing so. Although the focus has been on expanding a specific framework, other approaches, such as Torrance Tests of Creative Thinking \citep{torrance_1966_tests}, could offer complementary insights if customized to AI-generated music assessment. The value of interdisciplinary approaches in addressing complex problems that lie at the intersection of the Arts and Sciences is one of the building blocks of this research. 
However, several challenges remain. Notably, this work does not prescribe a method for combining the components (e.g., M-surprise, M-novelty, M-value) and sub-components (e.g., PS-surprise, HC-surprise) into a unified measure of M-Creativity. While a simple weighted sum such as:
\begin{equation}
\mathrm{M\text{-}Creativity} = \alpha \cdot \mathrm{M\text{-}value} + \beta \cdot \mathrm{M\text{-}novelty} + \gamma \cdot \mathrm{M\text{-}surprise}
\end{equation}
\noindent might seem attractive, this approach oversimplifies the interdependence among components. For instance, surprise often co-occurs with novelty, as novel creations tend to deviate from learned expectations. Similarly, the Wundt Curve \citep{berlyne1970novelty, wundt1904principles} suggests that hedonic value (like M-value) peaks at a balance between predictability (Low M-surprise) and novelty (M-novelty), further reinforcing that these components interact non-linearly. This ambiguity in assigning weights reflects the inner tension between subjective musical appreciation and the optimistic desire for objective evaluation. To address this, perhaps machine learning techniques like preference learning \citep{furnkranz_2010_preference} could be employed to infer $\{\alpha, \beta, \gamma\}$ from human-annotated datasets. Yet, this would go against the goal of objectively quantifying M-Creativity, since it would introduce human subjectivity into M-Creativity’s equation.

Ultimately, it must be emphasized that the proposed metrics are heuristic approximations, not replacements for human judgment. They overlook a plurality of considerations, such as cultural and emotional subtleties. Deep down, whether AI music generation systems could be truly creative is, as pointed out by \citet{boden_2004_the}, a rather philosophical question, whose answer depends on the agreed-upon definition of generative AI Musical Creativity (M-Creativity). This underscores that any attempt to score M-Creativity inevitably circles back to unresolved philosophical debates about the nature of creativity itself.

Reflecting on this paper's research questions, Boden's framework \citep{boden_2004_the} proves highly relevant to measuring musical creativity in AI-generated music, even though the concepts of value, novelty, and surprise are difficult to map fully onto the musical domain, and additional components may be needed to enhance her schema. The musical creativity of generative AI systems can nonetheless be rigorously defined, as demonstrated by the adopted interdisciplinary approach; the proposed definition is by no means absolute, being shaped by a modern Western perspective and subject to variation across demographics and time. The introduced metrics (see Table~\ref{tab:mcreativity-multipage}) can be applied to a wide range of models and outputs, used selectively to evaluate specific aspects of M-Creativity or combined into a single score. While M-Creativity can be quantified relatively objectively---or at least some of its components can---not all aspects are directly measurable, and some, such as M-value, can hardly be quantified objectively.

There are, of course, many limitations to this approach. Music and creativity are, at this stage, too complex to fully encapsulate and require several assumptions, such as PME's tribalism. Despite the interdisciplinary approach, too little consolidated knowledge exists to claim with certainty that the proposed metrics are good measures of musical creativity. In addition, this research focuses only on value, novelty, and surprise as perceivable in the produced output. Other classic definitions of creativity include different requirements for a creative product \citep{guilford1967creativity,kharkhurin2014creativity,runco2012standard,runco2025updating}, consider additional perspectives \citep{rhodes1961analysis}, or arise from different cultures \citep{lubart1990creativity}. While this framework is not an absolute or objective truth, it provides a useful middle ground: it generalizes not because it fully captures musical creativity, but because, without some structured framework, there is no meaningful way to analyze, compare, or enhance the creativity of AI-generated music.

\section{Ethical Considerations}

Exploring conceptual spaces to create novel, valuable, and surprising musical artifacts requires, in practice, navigating potential dangers. If, for instance, you generate a piece of music with racist lyrics or sounds that are undesirable, such as a sudden high-pitched frequency spike that could cause discomfort or even physical distress to listeners, the system risks producing outputs that are harmful or offensive, rather than creatively meaningful. These concerns highlight the tension between open-ended search and controlled generation: while creativity thrives in open-ended exploration, completely unconstrained systems may generate unintended, problematic, or even dangerous output. Without appropriate control mechanisms, AI-driven music generation could lead to ethically questionable or culturally insensitive artifacts that do not align with human values \citep{ecoffet_2020_open}. In addition to this, although the aim of this research is to explore how AI-generation systems can serve as potent creative tools, the boundary between an AI-assistive role and full replacement of music-making tasks is blurry and ought to be closely monitored to avoid socially and economically undesirable outcomes. It is essential for artists to proactively take part in building Creative AI to ensure fairer systems, particularly since most models scrape music on the Internet, disregarding copyright laws, which historically change more slowly than the technology they are now required to consider \citep{sunray_2021_sounds}. Recently, movements of protests have positioned themselves as anti-protagonists, which is necessary \citep{glynn_2025_artists}.

\section{Conclusion}

This paper proposed a multi-dimensional quantitative framework to evaluate the musical creativity of generative artificial intelligence music generation systems by expanding Boden's model of \textit{value}, \textit{novelty}, and \textit{surprise} \citep{boden_2004_the} to the musical domain. It introduced original computational metrics to quantify M-Creativity (\textit{M-value}, \textit{M-novelty}, and \textit{M-surprise}), each grounded in interdisciplinary definitions drawing from musicology, psychology, computing, neuroscience, and other fields. The detailed framework includes both artifact-level and model-level evaluations, using probabilistic, perceptual, and structural techniques across melodic, harmonic, rhythmic, and timbral dimensions. Rather than seeking to summarize musical creativity through a single unifying metric, it proposes an ensemble of complementary metrics that capture different dimensions of creativity. This reflects the complexity of musical artifacts and the difficulty of defining meaningful multi-objective compositional criteria across their multiple and interdependent properties. These metrics offer a systematic method to assess and compare AI-generated music, moving beyond purely subjective evaluation and laying the groundwork for more creative and human-aligned generative AI systems.

Beyond evaluation, the proposed metrics could also be used as objectives for fine-tuning generative music models. Individual metrics, or selected combinations of metrics, could be incorporated into reward functions or auxiliary training objectives to encourage specific dimensions of musical creativity, such as novelty, surprise, or value. This would provide a way to explicitly optimize models for creative properties in addition to conventional objectives such as likelihood or reconstruction quality. Given the multidimensional nature of musical creativity, however, such approaches would require careful consideration of how different objectives are combined, rather than reducing creativity to a single optimization target.

\bibliography{biblio}

\end{document}